\documentclass[preprint,showpacs,preprintnumbers,amsmath,amssymb]{revtex4}

\usepackage{graphicx}
\usepackage{dcolumn}
\usepackage{bm}

\catcode`\@=11
\def\lsim{\mathrel{\mathpalette\@versim<}}
\def\gsim{\mathrel{\mathpalette\@versim>}}
\def\@versim#1#2{\vcenter{\offinterlineskip
\ialign{$\m@th#1\hfil##\hfil$\crcr#2\crcr\sim\crcr } }}
\catcode`\@=12

\newcommand{\nn}{\nonumber}
\newcommand{\be}{\begin{equation}}
\newcommand{\ee}{\end{equation}}
\newcommand{\bea}{\begin{eqnarray}}
\newcommand{\eea}{\end{eqnarray}}
\newcommand{\bean}{\begin{eqnarray*}}
\newcommand{\eean}{\end{eqnarray*}}
\newcommand{\bit}{\begin{itemize}}
\newcommand{\eit}{\end{itemize}}
\newcommand{\ben}{\begin{enumerate}}
\newcommand{\een}{\end{enumerate}}
\newcommand{\ag}{\alpha_g}

\newcommand{\Dsla}{D \hspace{-7pt}/\hspace{2pt}}
\newcommand{\psla}{p \hspace{-5pt}/\hspace{2pt}}

\begin{document}

\title{
Fixed point merger in the $SU(N)$ gauge beta functions
}
\author{Yuki Kusafuka}
\author{Haruhiko Terao}
\email{terao@asuka.phys.nara-wu.ac.jp}
\affiliation{
Department of Physics, Nara Women's University
Nara, 630-8506, Japan
}%
\date{\today}
\begin{abstract}
We discuss qualitative behavior of the SU(N) gauge beta functions in
QCD with many massless flavors. Non-perturbative beta functions can be obtained 
by extracting the renormalized trajectories in the exact renormalization
group framework. 
We examine the renormalization group equations for the general four-fermi
couplings as well as the gauge coupling obtained in a simple approximation scheme.
It is shown that the gauge beta function possesses not only an IR but
also a UV fixed point in the conformal window. These fixed points
merge with each other and disappear at the edge of the conformal window.
The scaling dimensions of the quark mass operator at the fixed points
are also shown.

\end{abstract}



\maketitle

\section{Introduction}


Realization of scale invariant quantum field theories in four dimensions
is a very attracting subject not only as a theoretical interest but also
for the gauge hierarchy problem of the standard model. 
It has been known for sometime that non-abelian gauge theories
with appropriate numbers of massless fermions can be scale invariant.
Explicitly, the IR attractive fixed point, called the Banks-Zaks (BZ) 
fixed point \cite{bz}, has been shown to realize in the beta function of
an $SU(N_c)$ gauge theory, as long as the flavor number $N_f$ is less than
but close to $11N_c/2$.  Then the fixed point appears in the weak coupling
region and perturbative analysis is reliable.
As the flavor number is reduced, the fixed point coupling becomes strong 
and enters into the non-perturbative region eventually.

The non-perturbative dynamics of QCD with many flavors has been
studies by analyzing the so-called ladder 
Dyson-Schwinger (DS) equations \cite{SDappel,SDyama}.
It has been known that the chiral symmetry is spontaneously
broken for the gauge couplings stronger than a certain critical 
value.
This implies that the IR fixed point does not exist beyond the critical
gauge coupling, since scale invariance is lost by the dynamical scale generation.
The chiral symmetry breaking has been also examined \cite{chiralRG} by means of the Wilson
renormalization group \cite{wk}, 
or more explicitly by solving the exact (functional)
renormalization group (ERG) equations \cite{wk,wh,po,we,mo}.
The analyses using the ERG equations have been also performed for the
chiral dynamics in the conformal window \cite{gjconf,tt,BGscaling}.

Thus it is expected that the lower boundary of the conformal window $N_f^{cr}$
is determined by spontaneous breakdown of the chiral symmetry.
The DS equations in the large N and ladder approximation
leads to $N_f^{cr}=4N_c$, when combined with the two-loop beta function \cite{SDappel}.
A recent ERG analysis combined with the four-loop gauge beta function also leads to 
$N_f^{cr}=10.0_{-0.7}^{+1.6}$ for the $SU(3)$ case \cite{gjconf}.
Lattice Monte-Calro simulations have been also performed \cite{oldlattice}
and the results indicate existence of the conformal phase and the  critical
flavor number.
The recent simulations have provided evidence that $8< N_f^{cr}  \leq 12$ for the
$SU(3)$ QCD, even though the case $N_f=12$ is controversial \cite{Applattice,newlattice}.

In this paper we consider how the fixed point disappears from the gauge 
beta function when the flavor number goes out of the conformal window.
The IR fixed point in the two-loop perturbative beta function remains even 
below $N_f^{cr}$ and seems to move away to infinity 
at a certain $N_f$ lower than $N_f^{cr}$.
Meanwhile chiral symmetry breaking at the strong coupling region
prohibits existence of the fixed point.
Then it seems that the beta functions in the conformal window
cannot be continued  smoothly to the beta function with a fewer flavor number
\footnote{
It is remarkable that the conformal window has been explicitly found for
the $N=1$ supersymmetric QCD.
However, chiral symmetry is supposed to be intact beyond the conformal
window. Thus dynamical aspect seems rather different from the non-supersymmetric
case. The gauge coupling at the IR fixed point may become infinitely strong
at the edge of the conformal window. We will discuss this problem in the
last section.}.

The unique possibility is that a UV fixed point exists as well in the
conformal window and merges with the IR fixed point at the edge of the
conformal window. Then the IR fixed point can disappear before
reaching the critical gauge coupling for chiral symmetry breaking.
Such mechanism has been discussed in Ref.~\cite{klss} and showed that
the BKT scaling, or the so-called Miransky scaling 
\cite{miranskyscaling,SDyama} for the conformal
gauge theories, realizes.
In this paper we will show, although in a certain scheme of approximation,
that the gauge beta function possesses a UV fixed point and
fixed point merger occurs in practice.

In order to see what happens near the lower boundary of the
conformal window, we need to define the gauge beta function in the
non-perturbative region.
Here we shall apply the ERG formalism.  Because the
RG equations can be obtained without recourse to the perturbative
expansion, although some approximation is inevitable to solve
them.
The Wilson RG describes the flows in the infinite dimensional coupling space
by integrating out the higher momentum modes scale by scale \cite{wk}.
For a renormalizable theory, the flows converge towards a line called
the renormalized  trajectory (RT) in the continuum limit.
The RT corresponds to the renormalized theory.
The beta function for the renormalized theory is nothing but the
scale transformation on the RT, and therefore is given in a 
non-perturbative way in principle \cite{po}.

Here it should be noted that the Wilson RG  offers us a 
very suitable framework in seeking for the fixed points as well as 
the phase structure. It is also a great advantage to find out the
anomalous dimensions at the fixed points directly from the RG equations \cite{chiralRG}.
It is also remarkable that the Wilson RG is applicable not only in
the broken phase but also in the symmetric phase. 
Contrary to these features, the DS equations cannot
be applied in the symmetric phase. Therefore the DS
equations cannot examine the IR dynamics around the IR fixed point,
while it is useful to evaluate the order parameters in the broken
phases \cite{SDappel,SDyama}.

In the explicit calculation, we will incorporate the general 
four-fermi operators \cite{RG4fermi,tt,am} allowed by the symmetries
into the effective Lagrangian, 
but truncate other higher dimensional operators. 
The essential point is to extend the RG equation for the gauge
coupling so as to include the higher order corrections via the 
four-fermi operators. 
More explicitly, we give the RG flow equation for the gauge coupling 
by adding the gauge invariant
corrections via the effective four-fermi interactions to the two-loop perturbative
beta function.
This dependence on the four-fermi coupling enables us to take non-perturbative
corrections in the gauge beta function.
We also discard all the gauge non-invariant corrections as a primitive
analysis.

Thus we can define the gauge beta function containing 
non-perturbative corrections through non-renormalizable effective
operators.
The RG flow connecting the UV fixed point and the IR fixed point
is found to be the RT representing the continuum limit of the
asymptotically safe theory.
Therefore the non-perturbative gauge beta function also
possesses a UV fixed point.
Meanwhile it will be also shown that this RT is not achieved by
perturbative expansion of the RG flow equations.
Furthermore The UV fixed point in the
non-perturbative gauge beta function is found to merge with the IR, or the BZ,
fixed point at the boundary of the conformal window. 

Recently several ansatz for the non-perturbative gauge beta function have been
proposed \cite{betafnansatz,at}.
In Ref.~\cite{at}, the fixed point merger is imposed as one of the ansatz.
Contrary to these approaches, we do not make an ansatz, but try to 
consider the continuum limit and the beta function in a  tractable
approximation scheme. 
Therefore we do not intend to seek for the analytic forms of the non-perturbative
beta functions. 

This paper is organized as follows.
In section~2, we will discuss a toy example of a scalar theory to show
how we can define the non-perturbative beta function by applying the ERG framework.
Then, in section~3 we derive the RG equations for the $SU(N_c)$ gauge theory with $N_f$
massless flavors in the above mentioned approximation scheme.
In section~4, we will show the numerical results obtained by solving the
RG flow equations.
First the fixed point structure and the anomalous dimensions of fermion mass operator
at the fixed points are shown. 
The perturbative analysis for the gauge beta function are also
examined and discussed.
The non-perturbative gauge beta functions and the fixed point merger
are shown in section~5. 
The last section is devoted to the conclusions and discussions.

\section{Non-perturbative beta functions given by the Wilson RG}

In this section, we would like to consider  how to define the
beta function in a non-perturbative way by using the ERG
equations.
We may define the Wilsonian effective action by integrating out the
higher frequency modes by introducing momentum cutoff.
Then all operators allowed by the symmetry are induced in the
effective action irrespectively of the renormalizablity. 
Then the effective action is represented by a point in an infinite 
dimensional parameter space (theory space).
The Wilson RG follows response of the Wilsonian effective action
under scale transformation and, therefore, defines a RG flow for
each theory in the theory space.

In general, the flows converge towards a special flow called the
renormalized trajectory (RT) in the infinite cutoff limit (continuum
limit). 
The renormalizable theories are described in terms of the renormalized
trajectories. 
Accordingly the beta functions of the renormalized parameters
are given by the scale transformation on the RT in the Wilson RG
framework.    
It is noted that the Wilson RG equations may be given without
perturabative expansion, although approximations such as operator
truncation must be performed.
Therefore the beta function defined on the RT contains
non-perturbative informations, partly though.
In the followings, let us consider these aspects explicitly by examining
the Wilson RG equations for a  massless scalar $\phi^4$ theory.
   
We truncate the Wilsonian effective action as 
\be
\Gamma_{\Lambda} = \int d^4 x
\frac{1}{2}(\partial_{\mu} \phi)^2 
- \frac{\lambda_4}{4!} \phi^4 -\frac{\lambda_6}{6!\Lambda^2} \phi^6,
\ee
where we discarded the mass parameter just for simplicity.
Then the theory space is two dimensional and is spanned by 
$(\lambda_4, \lambda_6)$.
Here we may use the Wetterich equation \cite{we} given in terms 
of $t=\ln \Lambda$ by
\be
\partial_t \Gamma_{\Lambda}[\phi]=
\frac{1}{2}\mbox{STr} \partial_t R_{\Lambda}
\left( \Gamma_{\Lambda}^{(2)}[\phi] + R_{\Lambda} \right)^{-1},
\label{wettericheq}
\ee
where  $\Gamma_{\Lambda}[\phi]$ stands for the cutoff effective action and
$\Gamma_{\Lambda}^{(2)}$ denotes the second functional derivative
with respect to the scalar field $\phi$.
The function $R_{\Lambda}(p)$ is the regulator function, which plays a role of
cutoff around $p^2 \simeq \Lambda^2$.
Then the RG flow equations are found to be \cite{po} 
\bea
\Lambda \frac{d \lambda_4}{d \Lambda} &=& a \lambda_4^2 - b \lambda_6, 
\label{phi4ERG1}\\
\Lambda \frac{d \lambda_6}{d \Lambda}  &=& 2 \lambda_6 - c \lambda_4^3 + 2d \lambda_4 \lambda_6,
\label{phi4ERG2}
\eea
where the coefficients are dependent on the choice of the cutoff functions $R_{\Lambda}$.
If we take the sharp cutoff limit, then the coefficients are given as
\be
a=9A, ~~b=10A,~~c=27A,~~d=\frac{45}{2}A,~~~\left(A=\frac{1}{4 \pi^2}\right).
\ee
In Fig.~1, the RG flows are shown by the dotted lines. 
It is seen that the flows converge rapidly
towards the red line, which shows the RT obtained in the
continuum limit \cite{po}. 
Once the RT is found, then irrelevant coupling $\lambda_6$
is given as a function in terms of the renormalized coupling
$\lambda_4$, or  $\lambda_6=\lambda_6^*(\lambda_4)$.
Therefore the beta function is given explicitly as
\be
\beta_4(\lambda_4) = a \lambda_4^2 - b \lambda_6^*(\lambda_4).
\label{phi4NPbeta}
\ee

\begin{figure}[htbp]
\begin{center}
\includegraphics[width=80mm]{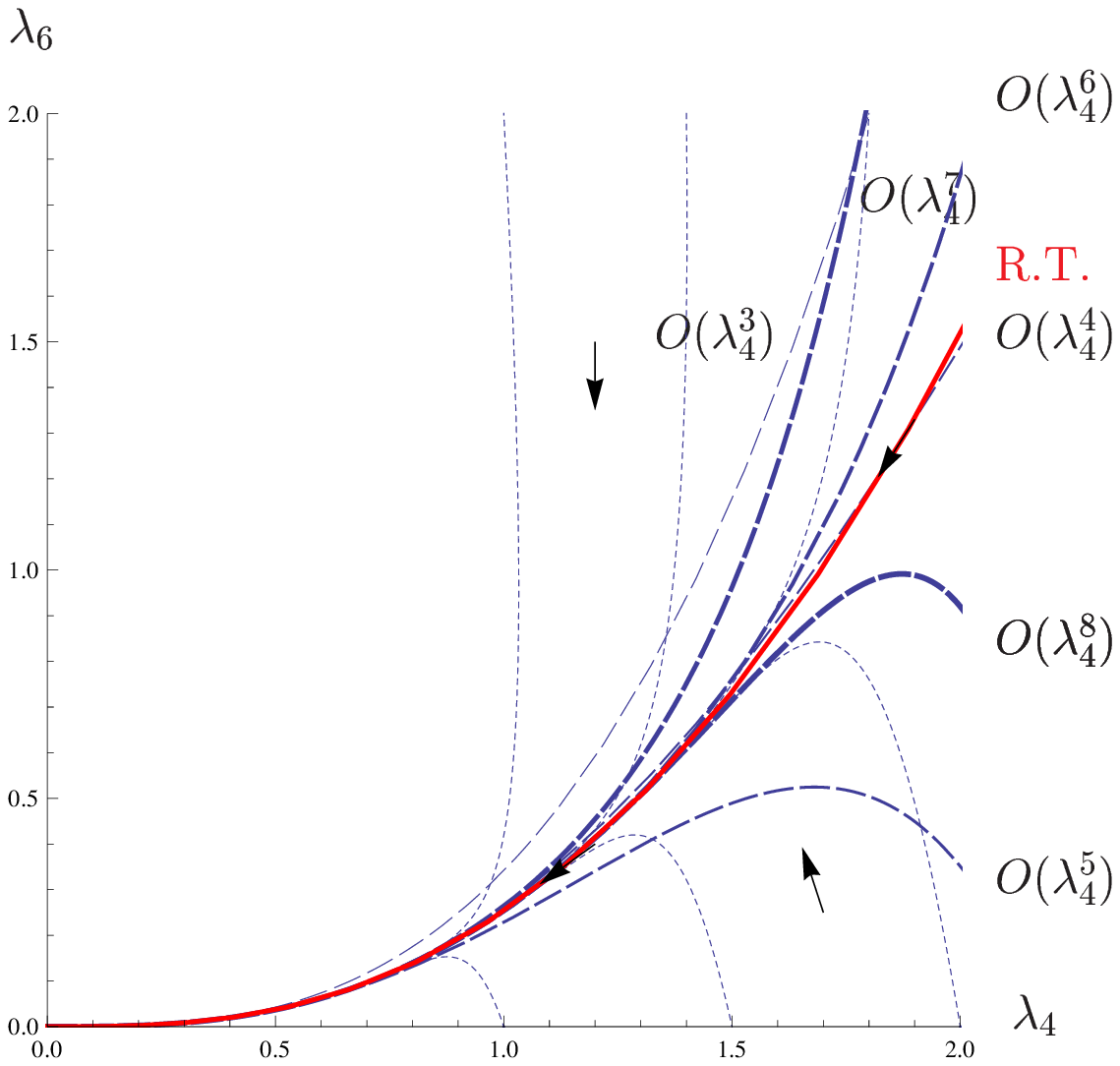}
\end{center}
\caption{RG flows obtained by solving Eq.~(\ref{phi4ERG1}) and (\ref{phi4ERG2}) are
shown by the dotted lines. The dashed lines represent the perturbative RTs obtained in
each order. The red line (R.T.) is the non-perturbative RT obtained numerically in the
continuum limit.}
\label{p4NPRT}
\end{figure}

Here we would stress that this beta function contains non-perturabative
corrections through the non-renormalizable coupling $\lambda_6$.
In order to see that, we compare this beta function with those obtained
by perturbative renormalization.
If we solve the RG equations by expanding the couplings $\lambda_i (i=4, 6)$
as
\be
\lambda_i=\lambda_i^{(0)} + A \lambda_i^{(1)} + A^2 \lambda_i^{(2)} + \cdots,
\ee
then the solutions are easily found order by order.

At the level of $O(A^0)$, the solutions are found to be 
\bea
\lambda_4^{(0)}(\Lambda) &=& \lambda_4^{(0)}(\Lambda_0), \\
\lambda_6^{(0)}(\Lambda) &=& (\Lambda/\Lambda_0)^2 \lambda_6^{(0)}(\Lambda_0).
\label{phi40th}
\eea
Here $\lambda_4^{(0)}(\Lambda)$ is fixed by imposing the renormalization
condition, while $\lambda_6^{(0)}(\Lambda)=0$ in the continuum limit, 
$\Lambda_0 \rightarrow \infty$.

Similarly we may proceed to $O(A^1)$. The RG equations
\bea
\Lambda \frac{d \lambda_4^{(1)}}{d \Lambda} &=& a  {\lambda_4^{(0)}}^2 - b  \lambda_6^{(0)}, \nn \\
\Lambda \frac{d \lambda_6^{(1)}}{d \Lambda} &=& 2 \lambda_6^{(1)} - c  {\lambda_4^{(0)}}^3 
+ 2d \lambda_4^{(0)} \lambda_6^{(0)},
\eea
give the solutions
\bea
\lambda_4^{(1)}(\Lambda) &=& a \ln \frac{\Lambda}{\Lambda_0}
{\lambda_4^{(0)}(\Lambda_0)}^2 
+ \lambda_4^{(1)}(\Lambda_0),
\label{phi41st1} \\
 \lambda_6^{(1)}(\Lambda) &=& 
\frac{c}{2}\left[1 -  \left(\frac{\Lambda}{\Lambda_0}\right)^2\right]
{\lambda_4^{(0)}(\Lambda_0)}^3+ \left(\frac{\Lambda}{\Lambda_0}\right)^2 
\lambda_6^{(1)}(\Lambda_0).
\label{phi41st2}
\eea
The logarithmic divergence in Eq.~(\ref{phi41st1}) is removed by the 1-loop counter term
$\lambda_4^{(1)}(\Lambda_0)$.
In the continuum limit, it is seen by noticing Eq.~(\ref{phi40th}) that 
\be
\lambda_6^{(1)}(\Lambda) ~~\rightarrow~~ \frac{c}{2}{\lambda_4^{(0)}(\Lambda)}^3.
\ee
This is nothing but the relation on the RT in the first level of perturbation.
In order to seek for solutions in the higher orders of $A$ , we may iterate the
similar calculations.
Then it is found that the RT is given as
\bea
\lambda_6^* &=&
 \frac{c}{2} \lambda_4^3 +\frac{c}{4}(3a-2d) \lambda_4^4  
- \frac{c}{8}(-12 a^2 + 3 b c + 14 a d - 4 d^2)\lambda_4^5 \nn \\
& &
+ \frac{c}{8}(30 a^3 - 18 a b c - 47 a^2 d + 10 b c d + 24 a d^2 - 
   4 d^3)\lambda_4^5 + \cdots.
\label{phi4pertRT}
\eea

In Fig.~1, the dashed lines represent the perturbative RTs obtained in
each order. The red line (R.T.) is non-perturbative RT obtained numerically in the
continuum limit.
It seems that the perturbative RTs approach to the non-perturbative RT obtained 
by solving RG flow equations directly. 
Thus the RT given by solving the ERG equations is found to contain
non-perturbative series.

In Fig.~2, the non-perturbative beta function defined by Eq.~(\ref{phi4NPbeta}) is given by
the red line (NP $\beta_4$). The dashed lines represent the perturbative beta 
functions obtained in each order.
It is seen that these beta functions converge to the non-perturbative beta function.
Thus the beta function containing non-perturabative informations can be
easily found by solving the Wilson RG equations.

\begin{figure}[htbp]
\begin{center}
\includegraphics[width=80mm]{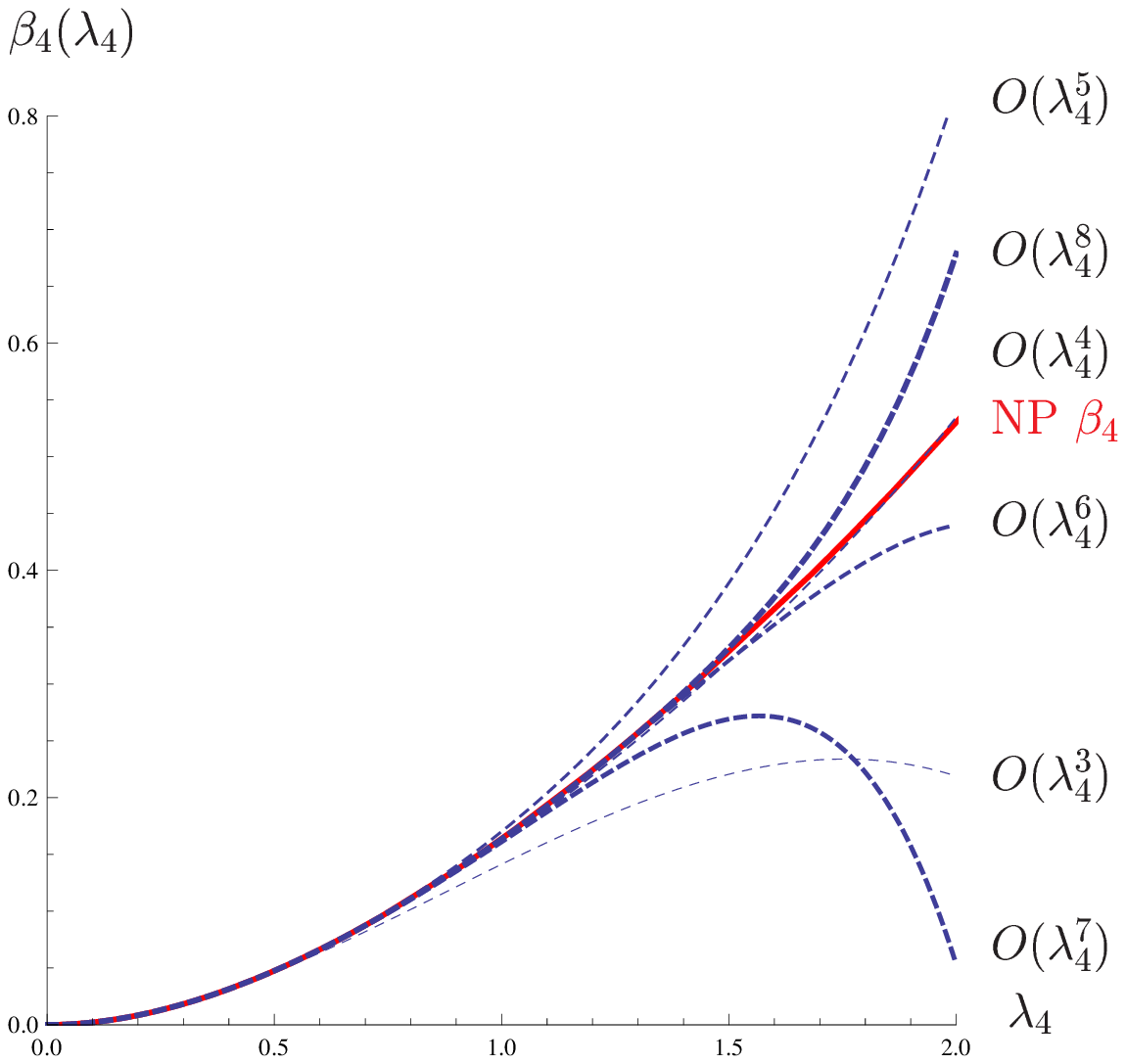}
\end{center}
\caption{The non-perturbative beta function defined by Eq.~(\ref{phi4NPbeta}) is given by
the red line (NP $\beta_4$). The dashed lines represent the perturbative beta 
functions obtained in each order.}
\label{NPbeta4}
\end{figure}

\section{Wilson RG for the $SU(N_c)$ QCD with $N_f$ flavors}

We consider the $SU(N_c)$ gauge theory with $N_f$ vector-like
massless fermions.
The theory satisfies the global chiral symmetry 
$SU(N_f)_L \times SU(N_f)_R$ as well as the $U(1)_V$ and the parity
symmetry interchanging the
left-handed and the right-handed fermions.
Therefore the Wilsonian effective action $\Gamma_{\Lambda}$ also should maintain
these symmetries in addition to the $SU(N_c)$ gauge symmetry.

It has been found that the phase structure of chiral symmetry is
revealed by incorporating the four-fermi operators in the effective
Lagrangian. Traditionally the chiral symmetry breaking has been studied
by means of the DS equations.
However, The Wilson RG has the following advantageous features to the 
DS approach. First it is straightforward to find out the
phase boundary as well as the fixed points.
Second, it is rather easy to improve the level of approximation by
including more operators in the Wilsonian effective action.
In practice, it has been shown that the non-ladder corrections can be
treated equally with the ladder corrections in QED and the gauge 
dependence, which bothers us in the DS approaches, is remarkably
improved \cite{chiralRG}.
Third, it is naturally performed to incorporate running effect of the
gauge coupling in the Wilson RG framework, while one may apply the 
Higashijima approximation in the DS equations \cite{hig}.

Therefore, it is important to include four-fermi operators in order
to examine the chiral phase structure and the conformal window of 
the $SU(N_c)$ gauge theory.
In this paper we take only the four-fermi operators in the Wilsonian
effective action $\Gamma_{\Lambda}$, but discard all other higher
dimensional operators as the first study;
\be
\Gamma_{\Lambda} = \int d^4 x~\left(
- \frac{1}{4g^2}F^A_{\mu\nu} F^{A~\mu\nu}
+ \bar{\psi}_f i \Dsla \psi^f + {\cal L}_{4f}
\right),
\ee
where the color index $a$ of massless fermions represented as
$\psi^{af}, (a=1, \cdots, N_c, f=1, \cdots, N_f)$ is suppressed
and ${\cal L}_{4f}$ denotes the four-fermi operators.
Strictly speaking, gauge variant operators are also
induced through the RG evolution due to momentum cutoff.
Therefore we should include these operators in the effective action
as well. However, we choose an easy way by discarding all gauge
variant corrections.
We shall come back this point later.

The four-fermi operators must be invariant under the $SU(N_c)$ gauge 
transformation, the $SU(N_f)_L \times SU(N_f)_R$ chiral transformation
and the parity.
We may write down various invariant operators, however some of them
are transformed to each other by the Fierz transformation.
It is shown in Appendix that there are 4 independent four-fermi
operators \cite{RG4fermi,tt,am}, which are given, for example, by 
\footnote{
We have chosen so that the Nambu Jona-Lasinio type four-fermi operator 
is one of the independent basis and the fermion bi-linears are always color
singlets \cite{tt}. This basis is slightly different from those given
in Ref.~\cite{RG4fermi}.
}
\be
{\cal L}_{4f} = 
\frac{G_S}{\Lambda^2} {\cal O}_S +
\frac{G_V}{\Lambda^2} {\cal O}_V +
\frac{G_{V1}}{\Lambda^2} {\cal O}_{V1} +
\frac{G_{V2}}{\Lambda^2} {\cal O}_{V2},
\label{4fefflag}
\ee
where
\bea
{\cal O}_S&=&
2 \bar{L}_i R^j \bar{R}_j L^i
= \frac{1}{2}\left[
\bar{\psi}_i \psi^j\bar{\psi}_j \psi^i
- \bar{\psi}_i \gamma_5 \psi^j \bar{\psi}_j \gamma_5 \psi^i 
\right], 
\label{4fos}\\
{\cal O}_V&=&
\bar{L}_i \gamma^{\mu} L^j \bar{L}_j \gamma_{\mu} L^i + (L \leftrightarrow R) 
= \frac{1}{2}\left[
\bar{\psi}_i \gamma^{\mu}\psi^j
\bar{\psi}_j \gamma_{\mu}\psi^i
+
\bar{\psi}_i \gamma^{\mu}\gamma_5 \psi^j
\bar{\psi}_j \gamma_{\mu}\gamma_5 \psi^i
\right], 
\label{4fov}\\
{\cal O}_{V1}&=&
2 \bar{L}_i \gamma^{\mu} L^i \bar{R}_j \gamma_{\mu} R^j
= \frac{1}{2}\left[
(\bar{\psi}_i \gamma^{\mu} \psi^i)^2
-
(\bar{\psi}_i \gamma^{\mu}\gamma_5 \psi^i)^2
\right], 
\label{4fov1}\\
{\cal O}_{V2}&=&
(\bar{L}_i \gamma^{\mu} L^i)^2 + (L \leftrightarrow R)
= \frac{1}{2}\left[
(\bar{\psi}_i \gamma^{\mu} \psi^i)^2
+
(\bar{\psi}_i \gamma^{\mu}\gamma_5 \psi^i)^2
\right]. 
\label{4fov2}
\eea
Here the color indices of fermions are always contracted and $i, j$
denotes the flavor indices. The left(right)-handed fermions are
represented by $L^{ai}=\psi^{ai}_L ~(R^{ai}=\psi^{ai}_R)$. 

The first operator ${\cal O}_S$ plays an essential role for
chiral symmetry breaking \cite{chiralRG}.
The coupling $G_S$ grows rapidly in the broken phase and diverges,
which indicates chiral symmetry breaking of
\be
\langle \bar{\psi}_i \psi^j \rangle = M^3 \delta_i^j.
\ee  
This non-trivial order parameter is expected just like in the  mean field approximation
of the Nambu-Jona Lasinio (NJL) model,
and then the flavor symmetry $SU(N_f)_L \times SU(N_f)_R$ is broken to
$SU(N_f)_V$. 

Now we shall derive the ERG equations for this system.
By expanding the Wetterich equation by operators, we may obtain
the RG equations for various effective couplings.
The trace form of the quantum corrections in the ERG equation given
by Eq.~(\ref{wettericheq}) implies that they
are just one-loop ones, although the vertices include non-renormalizable
interactions.
Therefore we may calculate the RG equations for the effective couplings
by evaluating various one-loop diagrams with shell integration
for the loop momentum.

{}For the four-fermi couplings, the one-loop diagrams are represented
as in Fig.~3, where four-fermi vertices represent any of the operators
given by Eqs.~(\ref{4fos}-\ref{4fov2}). 
The wavy line stands for the gauge propagator.
Thus we need to calculate rather many diagrams.
So we shall summarize details of the calculations in Appendix and
present only the results here.
\\

\begin{figure}[htb]
\begin{center}
\includegraphics[width=90mm]{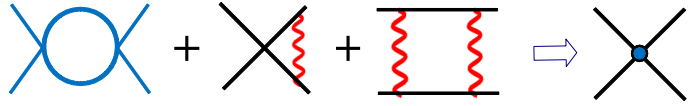}
\end{center}
\caption{The RG corrections for the effective 4-fermi couplings are
illustrated diagrammatically. The bold and wavy lines represent propagators
of the chiral fermions and the gauge bosons respectively. }
\end{figure}

We use the Landau gauge propagator and evaluate the loop integrations
in the sharp cutoff limit. We also use $g_i = G_i/4\pi^2 (i=S, V, V1, V2)$ 
and $\alpha_g=g^2/(4\pi)^2$.
Then the resultant RG equations are found to be
\bea
\Lambda \frac{d g_S}{d \Lambda}&=&
2 g_S - 2 N_c g_S^2 + 2N_f g_S g_V + 6 g_S g_{V1}
+2g_S g_{V2} \nn \\
& &
- 12 C_2(F)g_S \alpha_g + 12g_{V1}\alpha_g
- \frac{3}{2}\left(
3N_c - \frac{4}{N_c}-\frac{1}{N_c^2}
\right)\alpha_g^2, 
\label{gsfloweq} \\
\Lambda \frac{d g_V}{d \Lambda}&=&
2 g_V + (N_f/4) g_S^2 + (N_c+N_f) g_V^2 - 6 g_V g_{V2} \nn \\
& & 
- \frac{6}{N_c}(g_V + g_{V2})\alpha_g 
- \frac{3}{4}\left(
N_c - \frac{8}{N_c}+\frac{3}{N_c^2}
\right)\alpha_g^2, 
\label{gvfloweq}\\
\Lambda \frac{d g_{V1}}{d \Lambda}&=&
2 g_{V1} - (1/4) g_S^2 -g_Sg_V- 3 g_{V1}^2 - N_f g_S g_{V2}
+2(N_c+N_f)g_V g_{V1} \nn \\
& &+2(N_cN_f+1)g_{V1} g_{V2} 
+ \frac{6}{N_c}g_{V1}\alpha_g 
+ \frac{3}{4}\left(
1+\frac{3}{N_c^2}
\right)\alpha_g^2, 
\label{gv1floweq}
\eea
\bea
\Lambda \frac{d g_{V2}}{d \Lambda}&=&
2 g_{V2} - 3 g_V^2 - N_c N_f g_{V1}^2 + (N_c N_f-2) g_{V2}^2
- N_f g_S g_{V1} \nn \\
& &+2(N_cN_f+1)g_{V} g_{V2} 
+ 6 (g_{V}+g_{V2}) \alpha_g 
- \frac{3}{4}\left(
3+\frac{1}{N_c^2}
\right)\alpha_g^2,
\label{gv2floweq}
\eea
where $C_2(F)$ denotes the quadratic Casimir of color reprerentation of
the fermions and is given explicitly by $(N_c^2-1)/2N_c$.
In deriving these equations we do not make further approximations,
and therefore valid for any $N_c$ and $N_f$.

We need to solve these differential equations coupled with the
RG equation for the gauge coupling. Therefore the RG flows are
given in the five dimensional parameter space and
the flow diagram becomes rather complicated. 
Since our present purpose is qualitative understanding of the non-perturbative
 RT and the beta function,
let us coincide
the RG equations in the large $N_c$ and $N_f$ limit. 
This limit is taken by rescaling as
\bea
N_c g_{S(V)} &\rightarrow& g_{S(V)}, \\
N_c^2 g_{V1(V2)} &\rightarrow& g_{V1(V2)}, \\
N_c \alpha_g &\rightarrow& \alpha_g, 
\eea
with keeping the ratio $r=N_f/N_c$.
Then it is seen that the first two equations, (\ref{gsfloweq}) and (\ref{gvfloweq})
are reduced to  \cite{tt}
\bea
\Lambda \frac{d g_S}{d \Lambda}&=&
2 g_S - 2 g_S^2 + 2r g_S g_V - 6g_S \alpha_g
-\frac{9}{2}\alpha_g^2,
 \label{largeNgsfloweq} \\
\Lambda \frac{d g_V}{d \Lambda}&=&
2 g_V + \frac{r}{4} g_S^2 + (1+r) g_V^2 
-\frac{3}{4}\alpha_g^2.
\label{largeNgvfloweq} 
\eea
It is noted that four-fermi couplings $g_{V1}$ and $g_{V2}$ decouple
from the above equations. 
Therefore we may solve only three equations for the 
couplings $g_S, g_V$ and $\alpha_g$ in the large $N_c$ and $N_f$ limit.

Next we consider the RG equation for the gauge coupling.
The 2-loop beta function for the $SU(N_c)$ QCD is given by 
\be
\beta_g^{[2]}\equiv \mu \frac{d\alpha_g}{d \mu}
= -2 b_0 \alpha_g^2 - 2 b_1 \alpha_g^3,
\label{QCD2loopbeta}
\ee
where $\alpha_g= g^2/(4\pi)^2$ and
\bea
b_0&=& \frac{11}{3}N_c - \frac{2}{3}N_f, \\
b_1&=& \frac{34}{3}N_c^2 - N_f \left(
\frac{N_c^2-1}{N_c} + \frac{10}{3}N_c
\right).
\eea
Here we intend to incorporate non-perturbative corrections into
the gauge beta function in the same way performed for the scalar field
theory in section~2. 
It is thought that it is important to find  dependence on the
four-fermi couplings in the RG flow equation
for the gauge coupling.
Then some non-perturbative corrections are taken in the gauge
beta function through the effective four-fermi couplings.

However, it is not  straightforward to apply the RG flow equations for the
gauge theories. One problem is that the Wilson RG
does not respect manifest gauge invariance
\footnote{Though manifestly gauge invariant formulation of the ERG equations
has been investigated \cite{gaugeinvERG}, we do not apply it to
our problem for the simplicity.}.
Indeed, we may make use of the modified Slavnov-Taylor identities in order to
derive the RG flow equations for gauge invariant theories
\cite{modifiedSTI}.
Another problem is that we need to deal with fairly complicated flow equations
involving effective operators with higher derivatives in order to
reproduce even the two-loop beta function.
So we give up to apply the Wetterich equation for the gauge coupling directly 
and, in stead, 
derive the RG flow equation along the following considerations.

First, we substitute the part of the flow equation depending only on 
the gauge coupling  by the two-loop perturbative beta function given
in Eq.~(\ref{QCD2loopbeta}).
Then we add corrections depending on the four-fermi couplings as the
non-perturabative part of the beta function.

\begin{figure}[tbp]
\begin{center}
\includegraphics[width=40mm]{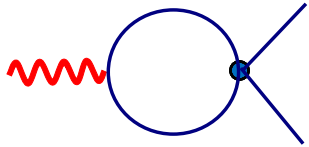}
\caption{The lowest order correction to the gauge vertex depending on the
effective four-fermi interactions.}
\label{vertexcorrection}
\end{center}
\end{figure}

We may think of one-loop corrections shown in Fig.~\ref{vertexcorrection}
inserting the four-fermi vertices as the non-perturbative
corrections.
The explicit calculation of the shell mode integration leads to
the extra contribution to the gauge beta function, 
\be
\delta \alpha_g = - r(g_S -2g_V) \alpha_g  \delta \ln \Lambda,
\ee
in the large $N_c$ and $N_f$ limit.
However, it is noted that these corrections should be forbidden in the
Landau gauge calculation, if the gauge invariance is protected.  
It may be also confirmed by the Slavnov-Taylor identity that the vertex correction
is indeed gauge variant \cite{RG4fermi}.
Therefore, we discard these corrections from our RG flow equation for
the gauge coupling.

\begin{figure}[tbp]
\begin{center}
\includegraphics[width=50mm]{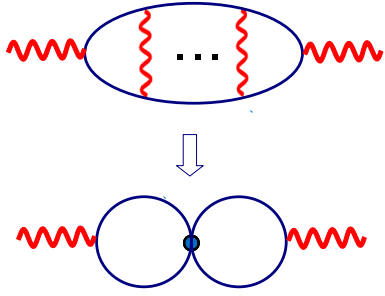}
\caption{The two-loop correction to the vacuum polarization of the gauge
field depending on the effective four-fermi interactions. }
\label{Z3corrections}
\end{center}
\end{figure}

However, we expect that there should exist gauge invariant corrections
to the gauge beta function through the effective four-fermi couplings. 
So let us consider the QED case. The higher order corrections to the
gauge beta function come from the vacuum polarization
diagrams. 
Especially we may focus on the ladder type diagrams shown
in Fig.~\ref{Z3corrections}
\footnote{Non-ladder diagrams are sub-leading in the large Nc limit. Planer
diagrams involving gauge boson vertices are not included in this
treatment. It is beyond our present scope to includes all planer diagrams
as well.}.
In the Wilson RG picture, a part of such contributions can be 
taken in by replacing the middle box diagram with the effective
four-fermi operators as shown in  Fig.~\ref{Z3corrections}. 

The box diagrams with fermions of the same chirality induce
the effective four-fermi operator given by
\bea
{\cal L}_{\rm eff} &\simeq&
\frac{G_V}{\Lambda^2}\left[
\bar{L}_i\gamma^{\mu} L^j~
\bar{L}_j \gamma_{\mu} T^A L^i
+(L \leftrightarrow R)
\right] \nn \\
&\simeq&
\frac{2G_V}{\Lambda^2}\left[
\bar{L}_i\gamma^{\mu}T^AL^i~
\bar{L}_j \gamma_{\mu} T^A L^j
+(L \leftrightarrow R)
\right],
\eea
where $\Lambda$ denotes the scale of the effective theory
and $G_V$ is the induced four-fermi coupling.
In the second line we performed the Fierz transformation
(see Appendix).
Then the four-fermi operator may be regarded as the current-current
interaction connected by a gauge propagator with momentum scale $\Lambda$.
Therefore such higher order contributions in the vacuum polarization diagrams
may be sum up by replacing the gauge boson exchange in the two-loop
diagram by 
$(g^2 + 2G_V ) D_{\mu \nu}$, where $D_{\mu\nu}$ denotes the gauge
boson propagator.
The two-loop vacuum polarization is calculated with dimensional regularization
in the gauge invariant way. 
Thus the vacuum polarization $\Pi_{\mu\nu}$
including the effective interaction is  found out to be
\be
\Pi_{\mu\nu}(q) = -N_f
\left[
\frac{4}{3}\alpha_g + 2 \alpha_g^2 + \alpha_g g_V
\right]\left(
q^2g_{\mu\nu} - q_{\mu}q_{\nu}
\right)\ln \Lambda,
\ee
where the epsilon pole in the dimensional regularization is
replaced by $\ln \Lambda$.

Consequently we may derive the RG flow equation for the gauge
coupling $\alpha_g$ as
\be
\Lambda\frac{d\alpha_g}{d \Lambda}=\beta_g^{[2]}
+ 2 r g_V \alpha_g^2, 
\label{QCDNPbeta}
\ee
where the rescaling for the large $N_c$ and $N_f$ limit has been performed
and the two-loop beta function $\beta_g^{[2]}$ is given by
\be
\beta_g^{[2]}=-\frac{2}{3}(11-2r)\alpha_g^2
-\frac{2}{3}(34-13r)\alpha_g^3,
\ee
in this limit. 
It is noted that the second term of Eq.~(\ref{QCDNPbeta}) corresponds to
$-b\lambda_6$ in Eq.~(\ref{phi4NPbeta}) defined for the scalar toy model, 
which contains non-perturbative informations.
We used the two-loop beta function $\beta_g^{[2]}$ as the perturbative
part,
since the non-perturbative part contains corrections only of more than the 
two-loop order. 
We may apply the three- or four-loop beta function \cite{4loopbeta,ps}
to the perturbative part as well, although the corrections may be overlapping
with the second term partly
\footnote{
The four-loop gauge beta function shows a UV fixed point \cite{4loopbeta, ps}.
However such a fixed is seen in general perturbative beta functions evaluated in
some higher orders, and it is a fake of perturbation. 
For example, the perturbative beta functions shown in Fig.~\ref{NPbeta4}
also show a UV fixed point in some orders. 
This fixed point should not be confused with the non-perturbative
fixed point shown in section~5.
Actually the fixed point gauge coupling is much smaller than the perturbative
fake couplings. 
}.
In this paper we perform explicit calculations of the RG flow equation
given by Eq.~(\ref{QCDNPbeta}) only.

\section{RG flows and the renormalized trajectories}

\begin{figure}[htbp]
\begin{center}
\includegraphics[width=60mm]{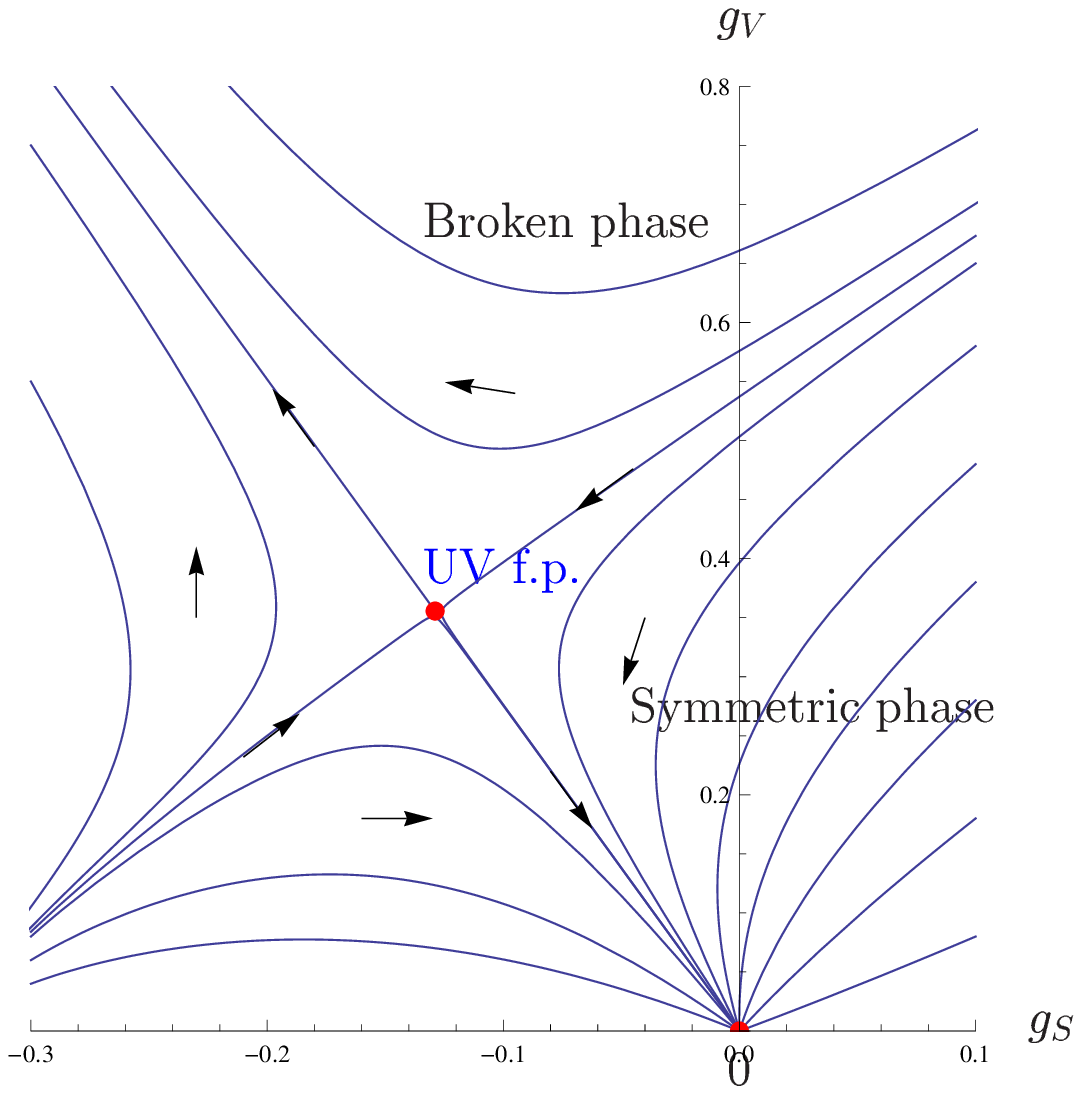}
\caption{Aspect of the RG flows of the pure four-fermi interactions
in the $g_S$-$g_V$ plane.}
\label{gsgvflows}
\end{center}
\end{figure}

In this section we show the RG flows obtained by solving numerically the RG flow
equations in the large $N_c$ and $N_f$ limit given by (\ref{largeNgsfloweq}),
(\ref{largeNgvfloweq}) and (\ref{QCDNPbeta}).
Before studying the gauge theories, let us examine the flows of the pure
four-fermi couplings by setting $\alpha_g=0$.
The RG flows in the $g_S$-$g_V$ plane are shown in Fig.~\ref{gsgvflows}.
It is seen that the RG flows are separated into two parts by the phase boundary.
The chiral symmetry is broken in the upper region, since
$g_S$ diverges towards the infrared direction.
On the phase boundary there is a UV fixed point, where $g_V$ takes
a negative value.
The IR fixed point is trivial. 
 
At the UV fixed point the four-fermi operators are recombined
into a relevant one and an irrelevant one. 
The relevant operator, 
\be
{\cal O}_{\rm rel}= \eta {\cal O}_S + {\cal O}_V,  
\ee
where the coefficient is $\eta = 2(-2-\sqrt{r^2-r+4})/r$.
The scaling dimension of this operator is found to be just
2 at the fixed point.
Though the relevant direction is given by a straight line as seen in 
Fig.~\ref{gsgvflows}, the line will be found to be curved in the presence of
the gauge interaction.

\begin{figure}[htbp]
\begin{center}
\includegraphics[width=80mm]{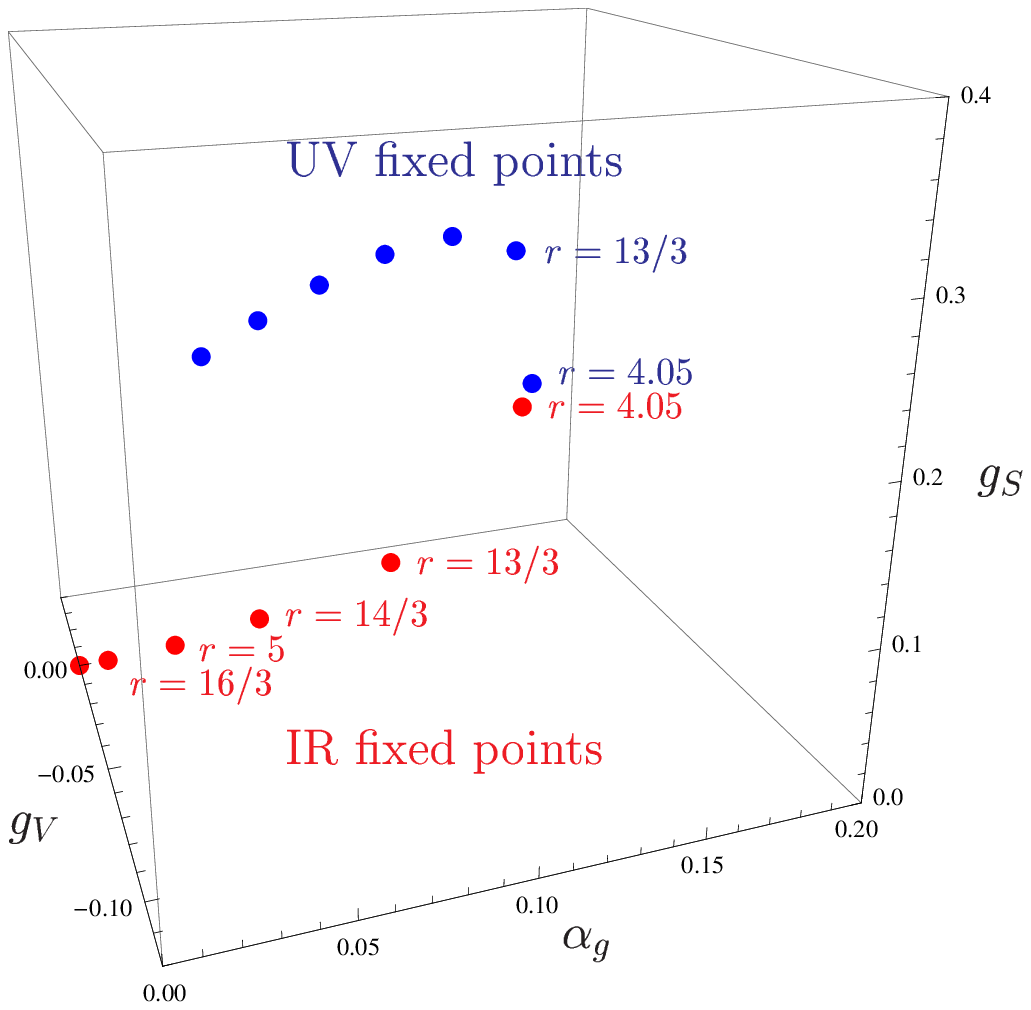}
\caption{UV and IR fixed points for $r=N_f/N_c=16/3, 15/3, 14/3, 13/3$
and $4.05$}
\label{figFP}
\end{center}
\end{figure}

Next we examine the RG flows of the gauge theories.
It is immediate to find the fixed points, since they are just given
by zero points of the RG flow equations (\ref{largeNgsfloweq}),
(\ref{largeNgvfloweq}) and (\ref{QCDNPbeta})
\footnote{
Although we may find many solutions as the roots, some of them
are thought to be fake.
We take only the fixed points linked with the IR
fixed point by the RG flow line.
}.
In Fig.~\ref{figFP}, the fixed points are shown in the 3 dimensional
coupling space for various numbers of $r=N_f/N_c$.
In the conformal window the IR fixed point carries non-trivial
couplings and a UV fixed point also appears.
As reducing the flavor number, these fixed point couplings
become strong. 
Eventually the UV fixed point and the IR fixed point merge with
each other and disappear around $r \simeq 4.05$.

\begin{figure}[htbp]
\begin{center}
\includegraphics[width=80mm]{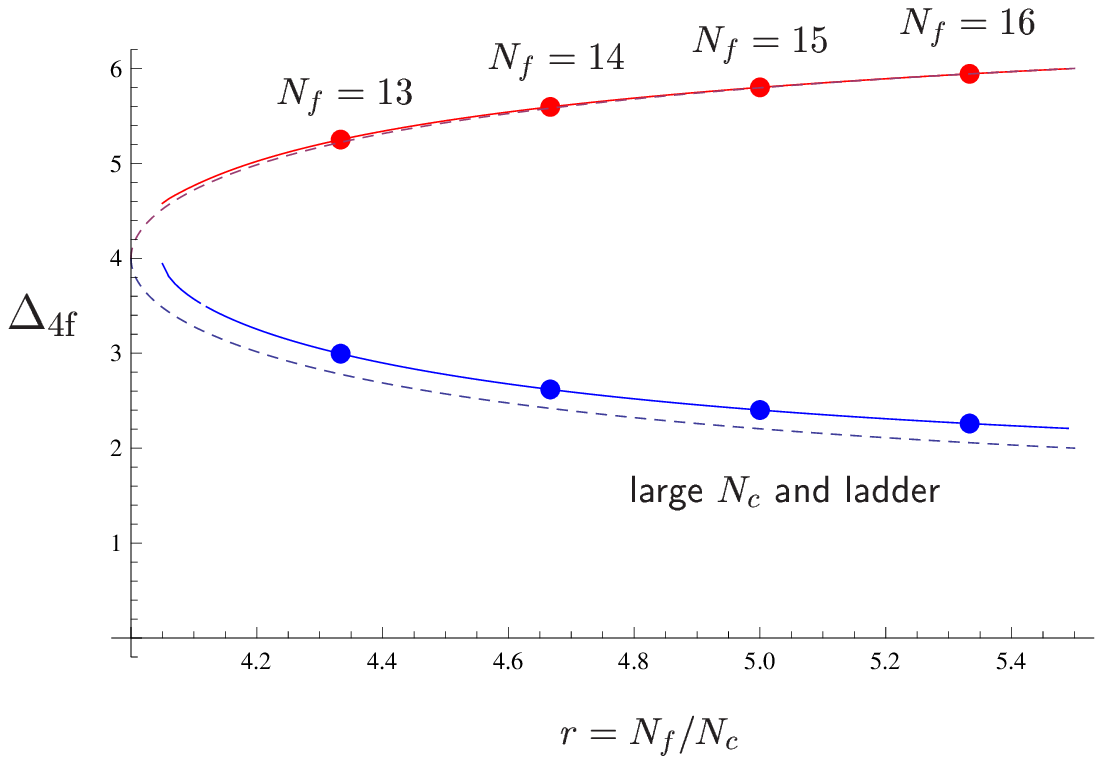}
\caption{Scaling dimensions of the most relevant four-fermion operator at the IR and
the UV fixed points in the conformal window.}
\label{figD4f}
\end{center}
\end{figure}

The scaling dimensions of the fermion operator at the vicinity of the
IR and the UV fixed points are important physical quantities
to be measured.
Especially, the eigenvalues of the RG flow equations linearized around
the fixed point are nothing but the scaling dimensions of
the four-fermi operators. 
We are interested in the most relevant dimension, since it shows
the (ir)relevance of the four-fermi operators at the fixed point.
Fig.~\ref{figD4f} shows the dimensions $\Delta_{\rm 4f}$ 
evaluated at the UV (blue) 
and the IR (red) fixed points in the cases of various flavor numbers \cite{gjconf,tt,BGscaling} .
The scaling dimension obtained in the large N and ladder approximation
is also shown for comparison. 
It is seen that the dimensions are almost same as those evaluated in 
the ladder approximation.
It is noted that the dimension should be 4 exactly at the edge of the
conformal window, since the four-fermi operator becomes marginal
by fixed point merger there.

\begin{figure}[htbp]
\begin{center}
\includegraphics[width=60mm]{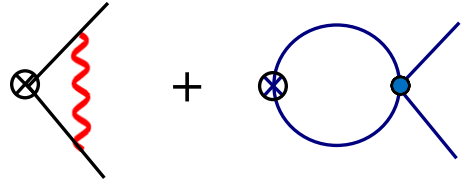}
\caption{Diagrammatic representation for the anomalous dimension of fermion mass operator.
The crossed vertex stands for the mass insertion.}
\label{gamma2f}
\end{center}
\end{figure}

\begin{figure}[htbp]
\begin{center}
\includegraphics[width=80mm]{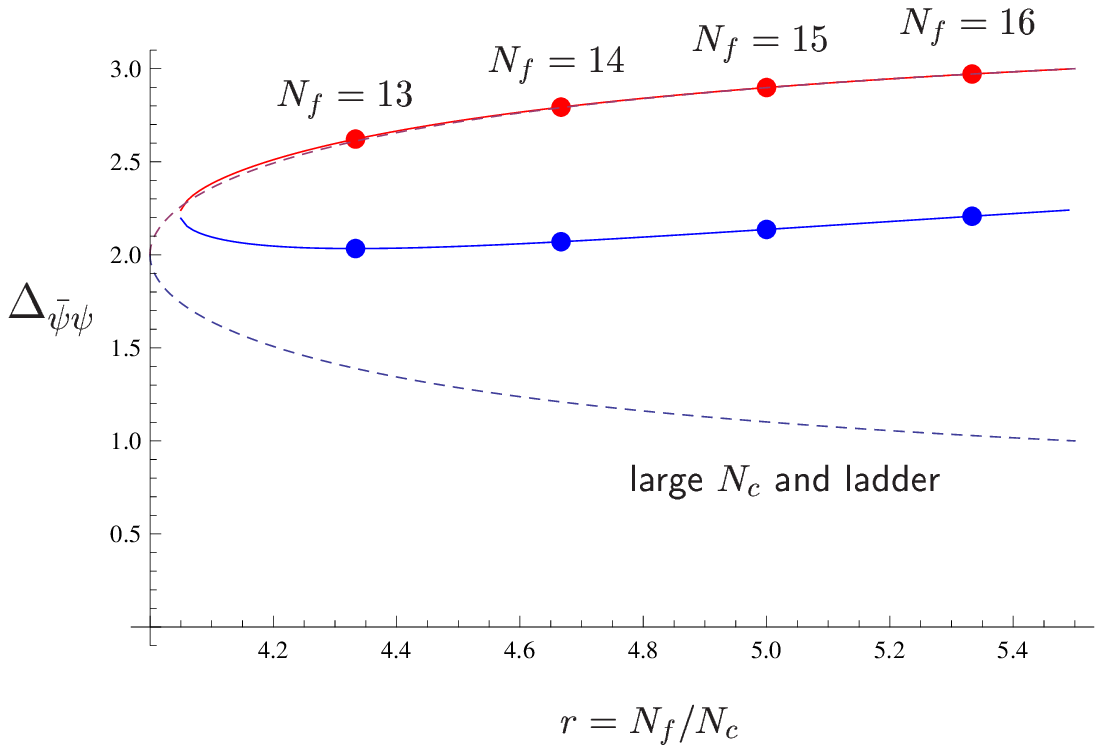}
\caption{Scaling dimensions of the fermion mass operator at the IR and the UV fixed
points in the conformal window.}
\label{figD2f}
\end{center}
\end{figure}

In the present scheme, the anomalous dimension of fermion mass is
simply represented by the diagrams given in Fig.~\ref{gamma2f} \cite{chiralRG,tt}.
The crossed vertex stands for the mass insertion.
It is immediate to evaluate these corrections at the fixed points
and the scaling dimension of the fermion mass operator
$\Delta_{\bar{\psi}\psi}$ 
are shown in Fig.~\ref{figD2f} for various flavor numbers
in the conformal window. 
We also present the results obtained in the large N and ladder
approximation for comparison. 
It is found that  the scaling dimensions at the
UV fixed points differ from the ladder results significantly, while 
there is not much differences at the IR fixed points \cite{tt}. 
It is noted that sum of the anomalous dimensions at the IR fixed point
and the UV fixed point is more than the value, $-2$, which is expected
in the large $N_c$ analysis \cite{kw}.
This is because the corrections taken in the RG flow equations contains not
only planer diagrams but also non-planer ones due to large $N_f$ contributions,
which influence the anomalous dimension significantly.  

\begin{figure}[htbp]
\begin{center}
\includegraphics[width=80mm]{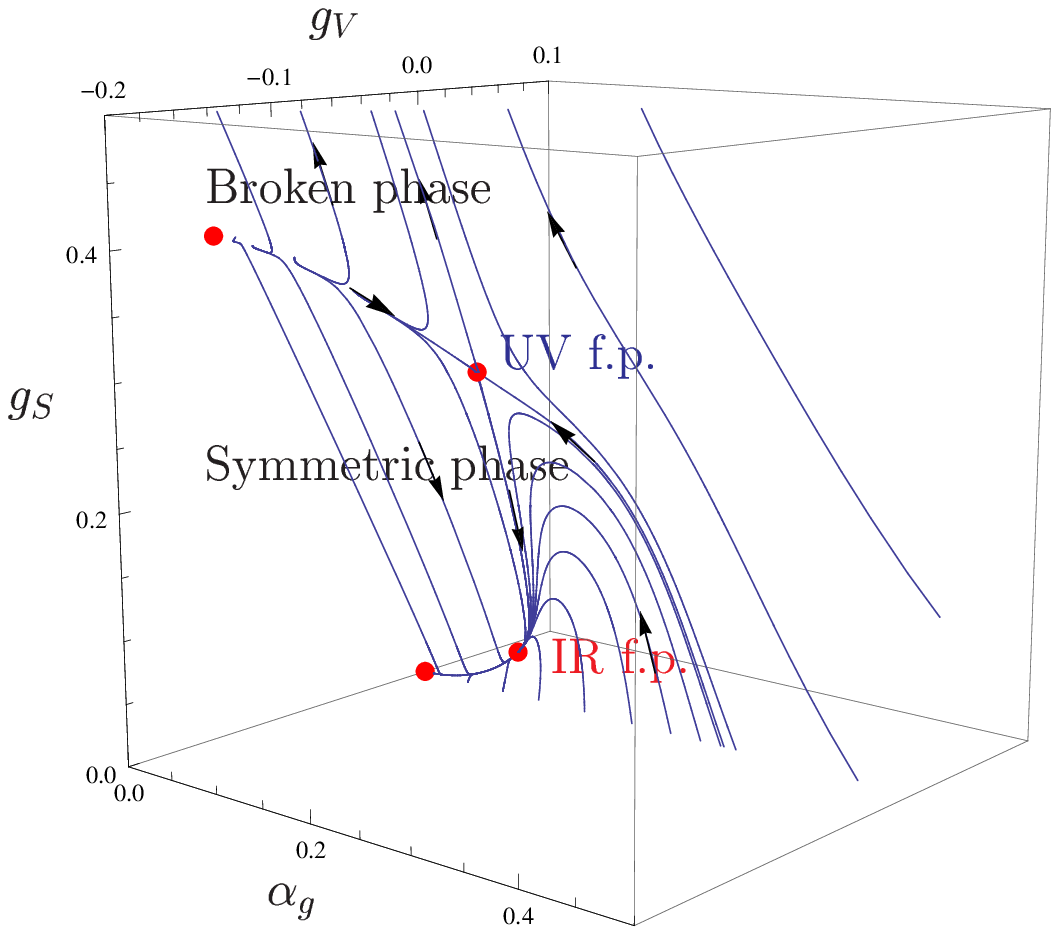}
\caption{Aspect of the RG flows in the 3 dimensional theory space in the case of
$N_f=13$. The fixed points are marked by red points. }
\label{flows3D13}
\end{center}
\end{figure}

\begin{figure}[htbp]
\begin{center}
\includegraphics[width=80mm]{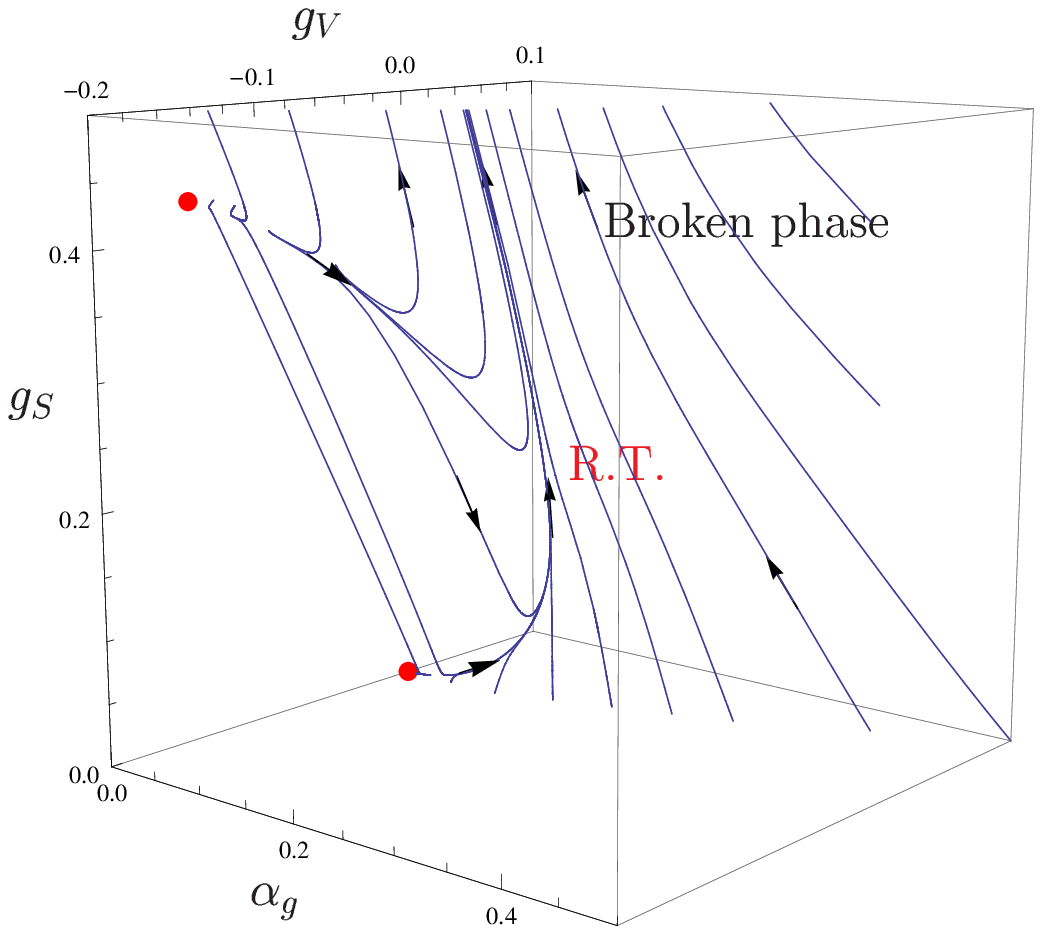}
\caption{Aspect of the RG flows in the case of $N_f=12$.}
\label{flows3D12}
\end{center}
\end{figure}

Next we consider change in behavior of the RG flows near the
lower edge of the conformal window.  
In Fig.~\ref{flows3D13} and Fig.~\ref{flows3D12}, the RG flows
of $SU(3)$ gauge theories are shown in the case with $N_f=13$ and
$N_f=12$ respectively.
The fixed points are shown by red points.
{}For $N_f=13$, the theory lies within the conformal window and
the UV and the IR fixed points with non-trivial gauge couplings
exist.
Meanwhile the theory with $N_f=12$ is slightly out of the conformal
window and these fixed points have annihilated in pair.

First let us discuss the RG flows of the theories in the conformal window.
As seen in Fig.~\ref{flows3D13},  there is the phase boundary and the chiral
symmetry is broken in the upper region.
The flows in the lower symmetric region run into the IR 
fixed point.
Then the flows converge towards a line connecting three fixed points;
the trivial fixed point, the IR fixed point and the UV fixed point.
This line is nothing but the renormalized trajectory.
The flow starting from the UV fixed point for the IR one  
corresponds to an asymptotically safe theory, while the flow approaching 
the IR fixed point from the trivial fixed point 
corresponds to an asymptotically free theory.
The both theories can be  defined as the continuum limit and, therefore,
renormalizable.

On the other hand, it is seen in Fig.~\ref{flows3D12} that
the non-trivial fixed points and the phase boundary disappear just below the
conformal window.
The four-fermi coupling $g_S$ diverges in all flows, therefore
the theories in the entire region belong to the broken phase.
The RG flows converge into a RT corresponding to the
the asymptotically free theory
\footnote{It has been shown that the gauged NJL models are also renormalizable
in some special cases \cite{gNJL}. It would be interesting problem to
clarify the asymptotically safe theories given by the RT emerging from
the UV fixed point of the pure 4-fermi interactions}.
The RT is unique and extended to infinity.
When the flavor number $N_f$ is lowered further, the RG flows show
the basically same structure and the RT remains in the QCD with
a small number of flavors. 
Now it is clearly seen from these two figures that the RT connecting
the trivial fixed point and the UV fixed point passing through
the IR fixed point for the theories in the conformal window is transformed
continuously to the asymptotically free RT below the conformal window.

\begin{figure}[htbp]
\begin{center}
\includegraphics[width=80mm]{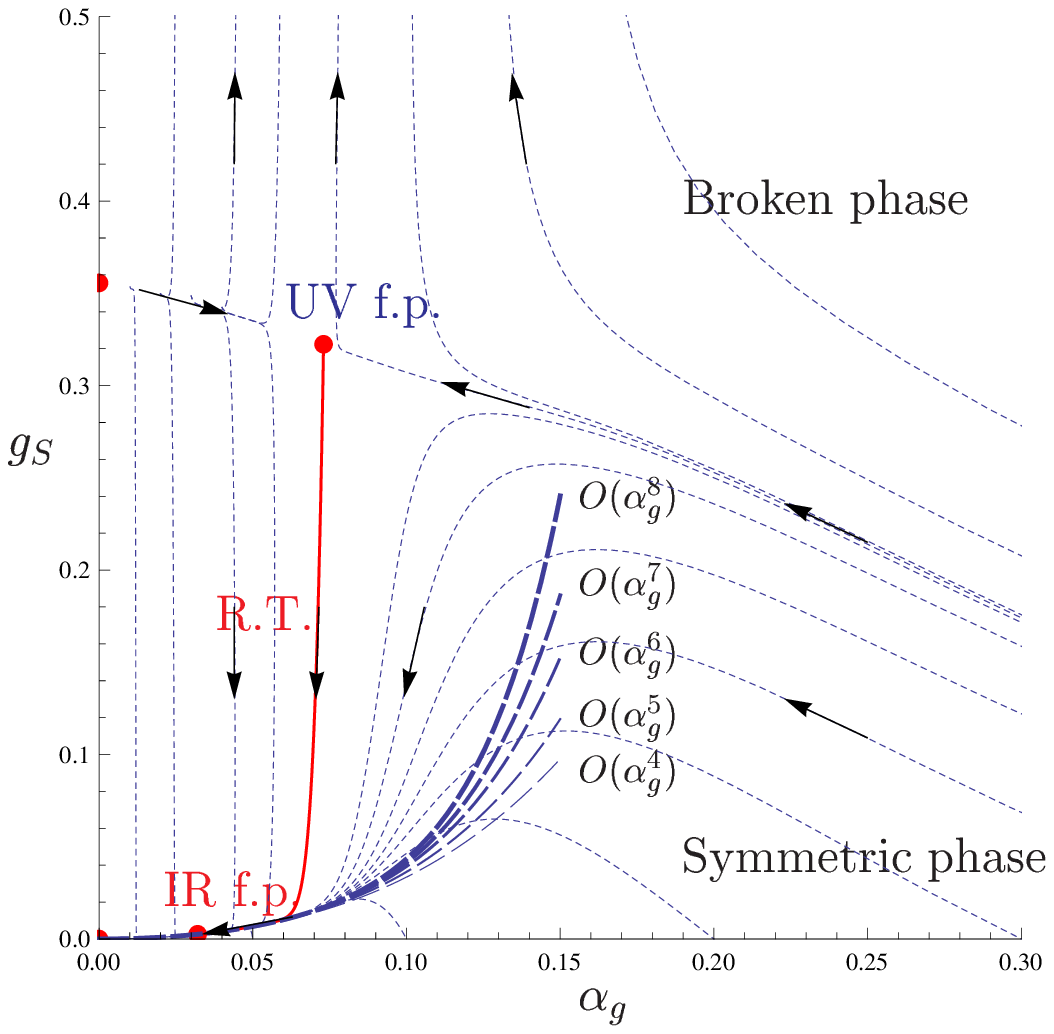}
\end{center}
\caption{The RG flows and the perturbative RTs obtained in various order calculations 
are shown in the $\alpha_g$-$g_S$ plane for $N_f=15$. The asymptotically safe RT 
is shown by a bold line.}
\label{pertRT15}
\end{figure}

However, when we look into the RG flows in the conformal window more
carefully, then  these show a somewhat curious feature as follows.
The RG flows obtained in the case of $N_f=15$, which is near upper bound
of the conformal window of the $SU(3)$ gauge theory, is shown in 
Fig.~\ref{pertRT15}.
The flows are projected on the $(\alpha_g, g_S)$ plane.
There is the UV fixed point as well as the IR fixed point and, therefore,
the flow connecting them is a RT,
while the flow connecting the trivial fixed point and the IR fixed point
is also another RT.

We note first that the RG flows in the symmetric phase 
converge into a distinct line from the RT, when the gauge coupling
approach the IR fixed point from the strong side.
The RT of the asymptotically free theory is unique.
Therefore, there seem to exist two different theories remaining in the
continuum limit. 

When the IR fixed point couplings are sufficiently weak, then the perturbative
analysis is reliable around the fixed point.
So we examine the RT by solving the RG flow equations given in (\ref{largeNgsfloweq}),
(\ref{largeNgvfloweq}) and (\ref{QCDNPbeta})
by perturbative expansion.
We may carry out the similar analysis performed for the scalar model
and find the following relations in the continuum limit,
\bea
g_S^* &=&
\frac{9}{4} \ag^2 -\frac{9}{4}(-3 + 2 b_0)\ag^3 
+\frac{9}{32}  (90 - 120 b_0 + 48 b_0^2 - 16 b_1 - 3 r) \ag^4 +\cdots, 
\label{QCDpertRTgs}\\
g_V^* &=&
\frac{3}{8}\ag^2 -\frac{3}{4} b_0 \ag^3 
+\frac{3}{128}  (-3 + 96 b_0^2 - 32 b_1 - 30 r)\ag^4 +\cdots.
\label{QCDpertRTgv}
\eea

In Fig.~\ref{pertRT15}, the RTs obtained in this manner up to the 8-th order
are also shown order by order.
It is seen that the asymptotically free RT is perfectly described by the
perturbative solutions.
However the series of these RTs converge to a
line distinct from the RT connecting the UV fixed point beyond the IR fixed point.
This shows that the theory described by the RT connecting
the UV fixed point cannot be obtained by perturbative
renormalization. Therefore we may say that the asymptotically safe theories
are truly non-perturbative in this sense.

While the perturbative RTs seem to converge around the
IR fixed point, they do not seem to converge for a larger gauge couplings.
Therefore it may be expected that the perturbative RT does not represent
the true continuum limit in the non-perturbative sense.
When the flavor number exceeds the conformal window, $N_f > (11/2)N_c$,
the theory becomes IR free
and behaves like QED
\footnote{In practice, the UV fixed point remains, when $N_f$ is slightly
more than $(11/2)N_c$. The RG flows similar to the QED case is obtained
after the UV fixed point is reduced to the pure four-fermi fixed point. }.
The continuum limit of 4 dimensional QED has been discussed in the
context of ERG and absence of the continuum limit is pointed
out \cite{gjQED}.

\begin{figure}[htbp]
\begin{center}
\includegraphics[width=80mm]{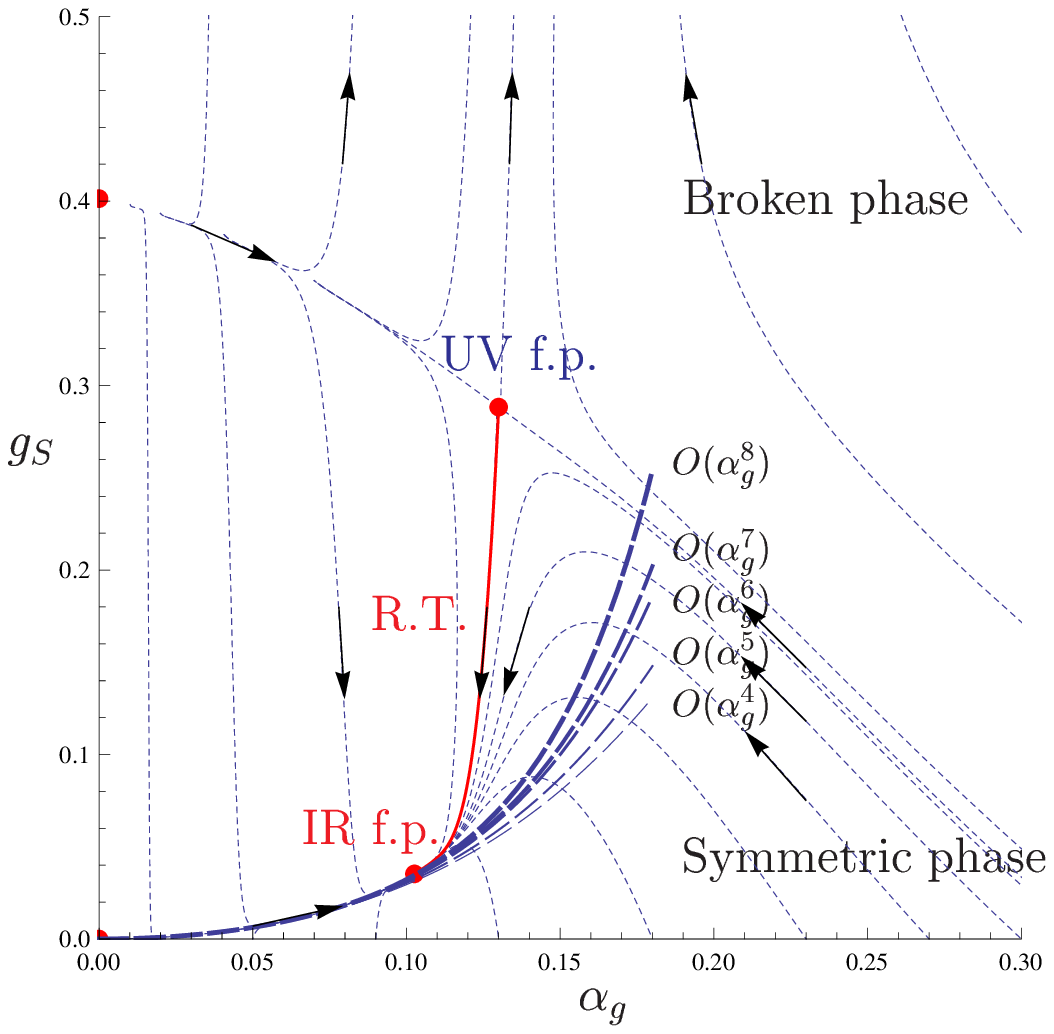}
\caption{RG flows and perturbative flows in the continuum limit projected
on the $\alpha_g$-$g_S$ plane for $N_f=13$}
\label{pertRT13}
\end{center}
\end{figure}

\begin{figure}[htbp]
\begin{center}
\includegraphics[width=80mm]{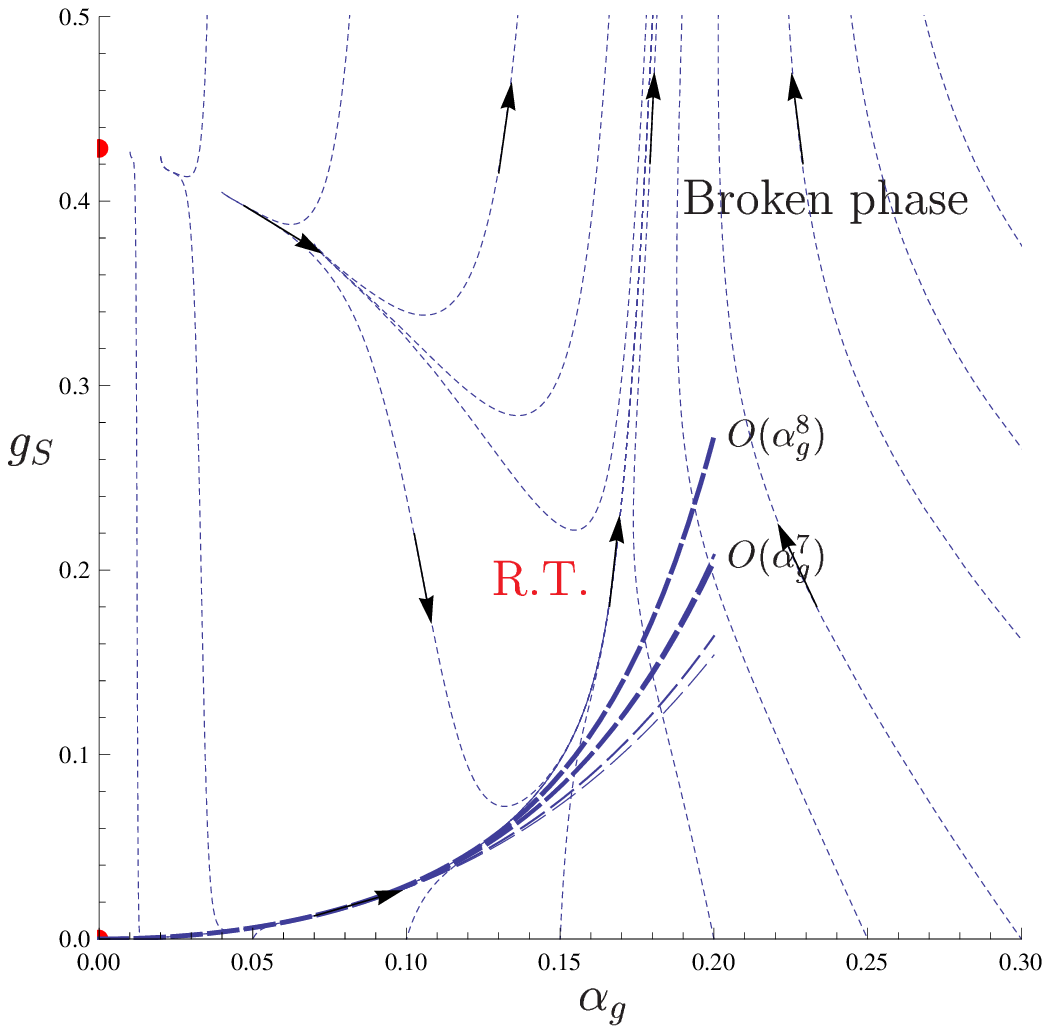}
\caption{RG flows and perturbative flows in the continuum limit projected
on the $\alpha_g$-$g_S$ plane for $N_f=12$}
\label{pertRT12}
\end{center}
\end{figure}

As the flavor number $N_f$ is reduced and the fixed point couplings become strong,
the perturbative RT also becomes close to the non-perturbative one.
In Fig.~\ref{pertRT13} and in Fig.~\ref{pertRT12}, the projected RG flows are shown
in the case of $N_f=13$ and $N_f=12$ the for $SU(3)$ QCD respectively.
There the RTs obtained by the perturbative expansion are also
shown by the dashed lines order by order.
Still the asymptotically free RT is found to be described well by the perturbative
series \footnote{
It has been also shown in Ref.~\cite{gg} that the IR fixed point
in QCD is perturbative in the entire conformal window. }.

It is seen in the figures that the perturbative RT gets closer to the RT passing through 
the UV fixed point for $N_f=13$ and becomes indistinguishable from the
asymptotically free RT for $N_f=12$.   
So it is supposed that the perturbative RT and the non-perturbative RT become
identical with each other out of the conformal window.
These statements are somewhat speculative, although we may naively expect
them by looking at the figures.
Of course, it is desirable to prove them mathematically, but it is beyond
our present purpose.

\section{Non-perturbative beta functions}

Now we discuss the non-perturbative gauge beta functions for the
many flavor QCD in and out of the conformal window.
As discussed in section 2, the non-perturbative beta function can be
given by the scale transformation on the RT.
It was also shown that the dependence on the non-renormalizable
couplings in the sense of the perturbative expansion is essential
to obtain the non-perturbative beta function.
Now the RG flow equation for the gauge coupling (\ref{QCDNPbeta}) 
contains a term dependent on
the four-fermi coupling $g_V$, which reflects the non-perturbative
corrections.

\begin{figure}[htbp]
\begin{center}
\includegraphics[width=80mm]{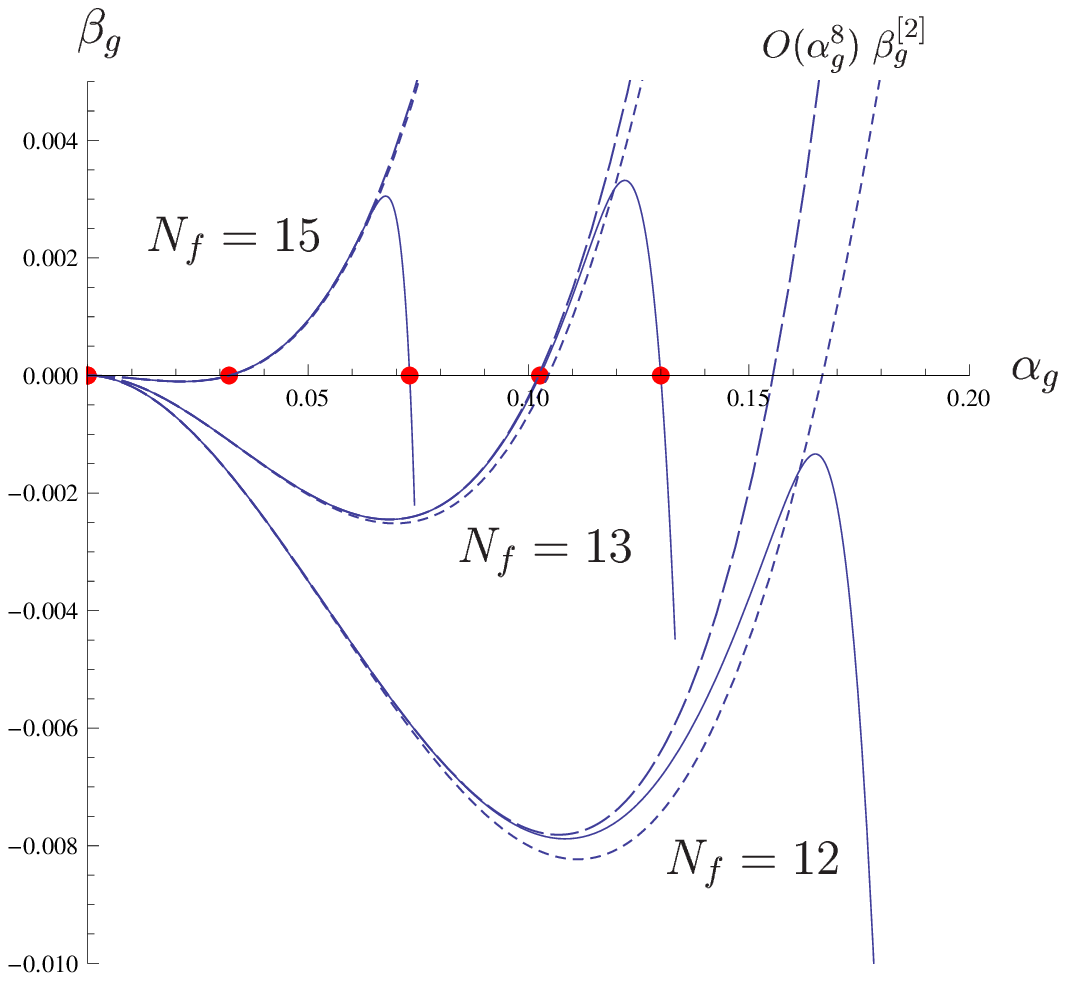}
\caption{The non-perturbative gauge beta beta functions are shown by the bold lines
in some cases of different flavor numbers.
The red points represent the fixed points. 
The perturbative beta functions and the two-loop beta functions
are also shown by dashed lines and the dotted lines respectively for comparison.}
\label{conformallost}
\end{center}
\end{figure}

First we examine the beta functions for the $SU(3)$ gauge theory
with $N_f=15$, which lies in the conformal window.
As we discussed, we can define the non-perturbative RT in addition to
the perturbative RT, which seems to be distinct in the conformal window.
The beta function depends on which RT we choose to define it.
We may define the non-perturbative beta function given by Eq.~(\ref{QCDNPbeta})
based on the non-perturbative RT, which transforms to the RTs out of the
conformal window smoothly. 
In Fig.~\ref{conformallost} the beta functions are shown by the bold lines.
The figure also presents how the non-perturbative
beta function changes it's form as the flavor number is reduced. 
It is seen that the gauge beta function possesses a UV fixed point in addition to
the IR fixed point in the conformal window and that these fixed points
merge with each other and disappear at the
edge of the conformal window.
Such a behavior of the beta function has been discussed in more general
contexts by Kaplan {\it et. al.} recently \cite{klss}.
Our analysis shows that the ``conformality lost'' may realize in the
gauge beta function of QCD with many flavors.

On the other hand, we may also define the perturbative beta functions by substituting
the solution given by Eq.~(\ref{QCDpertRTgv}) into the non-perturbative
beta function defined by Eq.~(\ref{QCDNPbeta}).
The beta functions are shown by the dashed lines in Fig.~\ref{conformallost}.
It is seen that the perturbative beta functions are almost same as the two-loop
beta functions, which are shown by the dotted lines.
Especially it is noted that no sign of the UV fixed point is seen for the perturbative
beta functions. 
Although the RT obtained by the perturbative calculation approach the
non-perturbative RT near the edge of the conformal window,
the beta functions are clearly different.
Thus we find that appearance of the UV fixed point in the gauge beta function is
purely due to non-perturbative effect.

One may wonder that disappearance of the fixed point
in the gauge beta function, or lost of the scale invariance, 
is due to fermion mass generation in the chiral symmetry broken phase. 
Furthermore, it may be thought that the bending behavior of the gauge beta function 
out of the conformal window is induced by fermion decoupling due to mass generation.
However, we may also make the following objections.
Suppose that the beta function possesses only the IR fixed point and
the gauge coupling ${\ag}_{\rm IR}$ is slightly less than the 
critical gauge coupling ${\ag}_{\rm cr}$ for the chiral symmetry breaking. 
According to the above picture, the beta function should bend in the region of 
$\ag > {\ag}_{\rm cr}$ by fermion decoupling. Then a UV fixed point
appears in the beta function inevitably. However this is forbidden in the
broken phase.
The gauge couplings at the fixed points ${\ag}_{\rm IR, UV}$
should be less than the critical coupling  ${\ag}_{\rm cr}$ and
this is exactly the case realized in our analysis.
Obviously the bending behavior of the non-perturbative beta function between 
the IR fixed point and the  
UV fixed point cannot be induced by fermion decoupling, since the
theory stays in the symmetric phase.

In our RG anlaysis chiral symmetry breaking occurs when the four-fermi couplings
goes to infinity. Therefore the scale of dynamically generated fermion mass should
be much less than the scale range of the RG flows in our analysis.
Accordingly the bending behavior of the non-perturbative beta function  out
of the conformal window is not caused by chiral symmetry breaking either.
It is an interesting problem to evaluate the dynamical scale in our RG
framework, though it is beyond our present scope.
Then  we may explicitly discuss the gauge beta function with fermion
decoupling effect and clarify the differences from the non-perturbative
beta function obtained in this paper.

\section{Conclusions and discussions}

In this paper we considered the gauge beta functions of the $SU(N_c)$
gauge theories in the conformal window. 
So far, dynamics of the chiral symmetry breaking near the conformal
window has been investigated by means not only of the DS
equations but also of the lattice MC simulations.
The phase structure and the IR/UV fixed points
in the conformal window have been already examined by the ERG approach
as well.
So far the analytic approaches adopted the perturbative gauge beta functions.
This paper presents the first study in order to understand
the non-perturbative aspect of the gauge beta function in and near the
conformal window.

It is shown explicitly that the non-perturbative beta function can be
evaluated without perturbative expansion in the Wilson RG.
Here, of course, non-perturbative does not mean exact, since we cannot help but
performing some approximations in the practical analysis of the RG flow equations.
We incorporated 4 independent four-fermi operators allowed by the symmetry
in the Wilsonian effective action, but discarded all other higher-dimensional operators.
Then the RG flow equations for these four-fermi couplings are derived
explicitly in the sharp cut-off scheme.
In practice, we analyzed the flow equations in the large $N_c$ and $N_f$ limit,
in order to discuss qualitative behavior of the gauge beta function.

We extended the RG flow equation for the gauge coupling so as to
include the higher order corrections through the four-fermi operators.
Explicitly we used two-loop beta function as the perturbative part
and added the non-perturbative corrections through the effective four-fermi 
coupling. 
This is the most important point in our analyses.
We discarded gauge variant corrections, which are
induced due to momentum cutoff, just to make the analysis simple.
Among all we neglected the vertex corrections with the four-fermi operators.

We obtained the  RG flows in the 3 dimensional coupling space by solving
the flow equations numerically.
In the conformal window, there exists a UV fixed point on the
critical surface of the chiral symmetry in addition to the IR fixed point. 
At the edge of the conformal window, these fixed points merge with each other
and disappear.
Apparently the flow connecting the UV fixed point and the IR fixed point
is a RT, which describes an asymptotically safe theory.
This RT is found to be transformed continuously to the RTs of the asymptotically free
theories out of the conformal window. 

On the other hand we may take the continuum limit of the perturbative
solutions of the RG flow equations.   
However it is found that the perturbative RTs do not converge to the 
asymptotically safe RT in the conformal window, although convergence
to the asymptotically free theory is remarkably good.
In other words, perturbative renormalization cannot describe
the asymptotically safe theory.
Moreover the series of the perturbative RTs itself does not seem to
converge at the strongly coupled region.
It is seen also that the perturbative RT coincides with the RT obtained
by RG flows in the theories below the conformal window.

We defined the non-perturbative gauge beta function, which gives
the scale transformation on the RT.
Then it is found that the gauge beta function possesses
not only the IR (BZ) fixed point but also a UV fixed point
and these fixed points merge
with each other at the edge of the conformal window.
Meanwhile the beta function evaluated by the perturbative analysis 
of the RG flow equations does not show the UV fixed point.
In addition, the perturbative beta function keeps an IR fixed point even below
the conformal window.
Thus it is said that the phenomenon of fixed point merger is 
owing to the non-perturbative
corrections taken in the gauge beta function.

Recently Kaplan et. al argued that the BKT type phase transition 
occurs in such situations in general \cite{klss}.
For gauge theories in the conformal window, this phase transition
has been known as the Miransky scaling \cite{miranskyscaling}.
In this paper we gave the explicit beta functions showing the ``conformality lost''
in the $SU(N)$ gauge theories. 

The order parameters of chiral symmetry breaking can be evaluated
by using the RG flow equations in the broken phase.  
It would be an interesting problem to verify the scaling properties
with respect to the flavor numbers
and compare with the BKT type scaling expected by the gauge beta function
\footnote{Recently the scaling of the chiral order parameter has been
considered in Ref.~\cite{BGscaling}. However the analysis is based on the
perturbative beta functions.}.
Further studies along this line of consideration will be reported elsewhere. 

Lastly let us make a comment on the beta function for the 
$N=1$ supersymmetric QCD.
It has been known that there is also the conformal window in the
supersymmetric QCD and the region is exactly
given by  $(3/2)N_c < N_f < 3N_c$ \cite{N=1SUSY}. 
It is remarkable that the anomalous dimension of the mass parameter
at the fixed point is related with the charge of R-symmetry carried
by the quark superfields and, therefore, is determined exactly.
Accordingly the anomalous dimension at the fixed point is unique.
This seems to imply that the fixed point is also unique, or that the
UV fixed point is absent.

Furthermore the N=1 supersymmetric QCD is known to satisfy 
the so-called exact beta function \cite{nsvz}, which is given in terms
of the anomalous dimension of the quark superfields.
If the anomalous dimension is a monotonic function of the gauge
coupling, then the beta function also has a unique fixed point.
We cannot know how the gauge coupling at the fixed point changes
in the conformal window, since we do not know the
anomalous dimesion as a function of gauge coupling except for the
perturbative calculation.
However uniqueness of the fixed point suggests that the fixed point
value moves to infinity in order to transform to the beta function
without a fixed point continuously.
If magnetic coupling of the dual gauge theory behaves like inverse
of the gauge coupling,
then the infinitely strong fixed point is thought to be allowed \cite{klss}. 
This feature seems quite contrasting with the non-perturbative
beta function discussed in this paper
\footnote{The essential difference in the gauge beta functions between QCD and
supersymmetric QCD has been also discussed in Ref.~\cite{gg}.}.
Thus extension of the present study to the supersymmetric cases will be also
quite interesting and remains for future studies.

\section*{Acknowledgements}
H.T thanks  Yoshio Kikukawa, Hiroshi Ohki,
Etusko Itoh, Tetsuya Onogi, Tim R. Morris  and Koichi Yamawaki 
for valuable discussions.   
H.T is supported in part by the Grants-in-Aid for 
Scientific Research (No.~20340053)
from the Ministry of Education, Science, Sports and 
Culture, Japan.

\appendix

\section{Invariant four-fermi operators}

In this section we show that the four-fermi operators given by
(\ref{4fos}-\ref{4fov2}) offer the independent basis of the invariant
four-fermi operators \cite{tt}.
The set of independent four-fermi operators has been also discussed
previously in Ref.~\cite{RG4fermi,am}, 
however, the operators given there are somewhat different from ours.
So, we make a brief explanation on the invariant four-fermi
operators and also present some useful formulae. 

The effective four-fermi interactions of massless $SU(N_c)$ gauge
theories should satisfy the global $SU(N_f)_L \times SU(N_f)_R$
chiral symmetry and the parity interchanging the left- and the 
right-handed  fermions in addition to the $SU(N_c)$ gauge symmetry. 
Here we use $a,b,\cdots$ for the color indices and $i,j,\cdots$ for the
flavor indices and represent components of a left- and a right-handed
chiral fermion by $\psi_L^{ai}=L^{ai}$ and $\psi_R^{ai}=R^{ai}$
respectively.
Then it is found that any invariant operator is given by a combination
of the following operators;
\bea
& & 
2\bar{L}_{ai}\gamma^{\mu}L^{ai}
\bar{R}_{bj}\gamma_{\mu}R^{bj}
= \frac{1}{2}\left[
(\bar{\psi}_{ai}\gamma^{\mu}\psi^{ai})^2
-(\bar{\psi}_{ai}\gamma^{\mu}\gamma_5\psi^{ai})^2
\right], 
\label{A1}\\
& &
2\bar{L}_{ai}\gamma^{\mu}L^{bi}
\bar{R}_{bj}\gamma_{\mu}R^{aj}
= \frac{1}{2}\left[
\bar{\psi}_{ai}\gamma^{\mu}\psi^{bi}
\bar{\psi}_{bj}\gamma_{\mu}\psi^{aj}
-\bar{\psi}_{ai}\gamma^{\mu}\gamma_5\psi^{bi}
\bar{\psi}_{bj}\gamma_{\mu}\gamma_5\psi^{aj}
\right], 
\label{A2}\\
& &
\bar{L}_{ai}\gamma^{\mu}L^{ai}
\bar{L}_{bj}\gamma_{\mu}L^{bj} + (L \leftrightarrow R)
= \frac{1}{2}\left[
(\bar{\psi}_{ai}\gamma^{\mu}\psi^{ai})^2
+
(\bar{\psi}_{ai}\gamma^{\mu}\gamma_5\psi^{ai})^2
\right], 
\label{A3}\\
& &
\bar{L}_{ai}\gamma^{\mu}L^{bi}
\bar{L}_{bj}\gamma_{\mu}L^{aj}+ (L \leftrightarrow R)
= \frac{1}{2}\left[
\bar{\psi}_{ai}\gamma^{\mu}\psi^{bi}
\bar{\psi}_{bj}\gamma_{\mu}\psi^{aj}
+
\bar{\psi}_{ai}\gamma^{\mu}\gamma_5\psi^{bi}
\bar{\psi}_{bj}\gamma_{\mu}\gamma_5\psi^{aj}
\right],
\label{A4}\\
& &
\bar{L}_{ai}\gamma^{\mu}L^{aj}
\bar{L}_{bj}\gamma_{\mu}L^{bi}+ (L \leftrightarrow R)
= \frac{1}{2}\left[
\bar{\psi}_{ai}\gamma^{\mu}\psi^{aj}
\bar{\psi}_{bj}\gamma_{\mu}\psi^{ai}
+
\bar{\psi}_{ai}\gamma^{\mu}\gamma_5\psi^{aj}
\bar{\psi}_{bj}\gamma_{\mu}\gamma_5\psi^{bi}
\right], 
\label{A5}\\
& &
\bar{L}_{ai}\gamma^{\mu}L^{bj}
\bar{L}_{bj}\gamma_{\mu}L^{ai}+ (L \leftrightarrow R)
= \frac{1}{2}\left[
\bar{\psi}_{ai}\gamma^{\mu}\psi^{bj}
\bar{\psi}_{bj}\gamma_{\mu}\psi^{ai}
+
\bar{\psi}_{ai}\gamma^{\mu}\gamma_5\psi^{bj}
\bar{\psi}_{bj}\gamma_{\mu}\gamma_5\psi^{ai}
\right].
\label{A6}
\eea

Then we may obtain following formulae of the Fierz transformation,
which are useful in calculations
of the RG flow equations for the four-fermi couplings;
\be
\bar{\psi}_1\gamma^{\mu}\psi_2
\bar{\psi}_3\gamma_{\mu}\psi_4
+
\bar{\psi}_1\gamma^{\mu}\gamma_5\psi_2
\bar{\psi}_3\gamma_{\mu}\gamma_5\psi_4 
=
\bar{\psi}_1\gamma^{\mu}\psi_4
\bar{\psi}_3\gamma_{\mu}\psi_2
+
\bar{\psi}_1\gamma^{\mu}\gamma_5\psi_4
\bar{\psi}_3\gamma_{\mu}\gamma_5\psi_2, 
\ee
and
\be
\bar{\psi}_1\psi_2
\bar{\psi}_3\psi_4
-
\bar{\psi}_1\gamma_5\psi_2
\bar{\psi}_3\gamma_5\psi_4 \\
=
-\frac{1}{2}\left[
\bar{\psi}_1\gamma^{\mu}\psi_4
\bar{\psi}_3\gamma_{\mu}\psi_2
-
\bar{\psi}_1\gamma^{\mu}\gamma_5\psi_4
\bar{\psi}_3\gamma_{\mu}\gamma_5\psi_2 
\right].
\ee
By using these identities the operators given in (\ref{A1}-\ref{A6}) are
rewritten into
\bea
{\rm (A1)} &=& 
-\left( \bar{\psi}_{ai}\psi^{bj} \bar{\psi}_{bj}\psi^{ai}
-\bar{\psi}_{ai}\gamma_5 \psi^{bj} \bar{\psi}_{bj}\gamma_5 \psi^{ai}\right) 
= -4 \bar{L}_{ai}R^{bj} \bar{R}_{bj}L^{ai}, \\
{\rm (A2)} &=& 
-\left( \bar{\psi}_{ai}\psi^{aj} \bar{\psi}_{bj}\psi^{bi}
-\bar{\psi}_{ai}\gamma_5 \psi^{aj} \bar{\psi}_{bj}\gamma_5 \psi^{bi}\right) 
= -4 \bar{L}_{ai}R^{aj} \bar{R}_{bj}L^{bi}, \\
{\rm (A3)} &=& 
\frac{1}{2} \left( \bar{\psi}_{ai} \gamma^{\mu} \psi^{bj} \bar{\psi}_{bj}\gamma_{\mu} \psi^{ai}
-\bar{\psi}_{ai}\gamma^{\mu}\gamma_5 \psi^{bj} \bar{\psi}_{bj}\gamma_{\mu}\gamma_5 \psi^{ai}
\right) 
= {\rm (A6)}, \\
{\rm (A4)} &=& 
\frac{1}{2} \left( \bar{\psi}_{ai} \gamma^{\mu} \psi^{aj} \bar{\psi}_{bj}\gamma_{\mu} \psi^{bi}
-\bar{\psi}_{ai}\gamma^{\mu}\gamma_5 \psi^{aj} \bar{\psi}_{bj}\gamma_{\mu}\gamma_5 \psi^{bi}
\right) 
=  {\rm (A5)}, 
\eea
Therefore it is seen that there are 4 independent operators.
We have selected the operators in which the color indices are contracted
with in the bi-linears among all.
Then the operators given by (\ref{4fos}-\ref{4fov2})
are chosen as the independent basis of the  invariant 4-fermi operators.
It is also noted that the operator ${\cal O}_S$, which is the NJL type four-fermi
operator, is found to be suitable to explore the dynamics of chiral symmetry breaking.  

The Fierz identities containing the color generator matrix $T^A$
are also given by using the completeness identity, 
\be
2 \sum_{A=1}^{{\rm dim}G}
\left(T^A\right)_d^a \left(T^A\right)_b^c
= \delta^a_b \delta^c_d - \frac{1}{N_c}
\delta^a_d \delta^c_b.
\label{groupid}
\ee
Then the formulae are given as follows;
\bea
& &
2 \sum_A~\bar{L}_i T^A \gamma^{\mu} L^i~
\bar{R}_j T^A \gamma_{\mu} R^j
= -{\cal O}_S - \frac{1}{2N_c}{\cal O}_{V1} \\
& &
\sum_A~\bar{L}_i T^A \gamma^{\mu} L^i~
\bar{L}_j T^A \gamma_{\mu} L^j +(L\leftrightarrow R)
= \frac{1}{2}{\cal O}_V - \frac{1}{2N_c}{\cal O}_{V2} \\
& &
\sum_A~\bar{L}_i T^A \gamma^{\mu} L^j~
\bar{L}_j T^A \gamma_{\mu} L^i +(L\leftrightarrow R)
= \frac{1}{2}{\cal O}_{V2} - \frac{1}{2N_c}{\cal O}_V
\eea
These identities are also useful in deriving the RG flow equations given
by Eq.~(\ref{gsfloweq}-\ref{gv2floweq}).

\section{RG flow equations for the four-fermi couplings}

In this section, we show a sketch of derivation of the RG flow equations
given by (\ref{gsfloweq}-\ref{gv2floweq}) in the sharp cut off limit.
Corrections in the Wetterich equation (\ref{wettericheq}) are
represented as one-loop diagrams with effective couplings.
Therefore the following diagrammatical derivation would be the easiest.
The momentum integration is restricted into an infinitesimally
small shell region by the derivative of the cutoff function
$\partial_t R_{\Lambda}$. 
In the sharp cut off limit we may use
\be
\int'dp \equiv \int\frac{d^4 p}{(2\pi)^4} \delta(|p|-\Lambda) \delta t,
\ee
where $t=\ln \Lambda$, and $p$ stands for the Euclidean momentum.

We need to evaluate corrections for the effective four-fermi couplings,
and, therefore, consider the one-loop diagrams with four external fermions.
However there are many diagrams to be concerned, and we shall show only
typical two kinds of diagrams among them here.

First we consider the diagrams using two effective four-fermi operators
${\cal O}_S$. There are some diagrams depending on the ways of contraction.
Here we consider only diagrams shown in Fig.~17.
The correction represented by the diagram (a) is explicitly given by 
\bea
{\rm (a)} &=&
-\left(\frac{2G_S}{\Lambda^2}\right)^2 N_c 
\int'dp ~{\rm tr} \left(
\frac{i \delta_k^i}{\psla} \frac{1+\gamma_5}{2} \frac{i\delta_j^l}{\psla}
\right) 
\bar{L}_i R^j \bar{R}_l L^k \nn \\
&=& 
\frac{N_c}{2\pi^2} G_S^2\delta t ~\frac{1}{\Lambda^2}{\cal O}_S.
\eea
Therefore this correction induces variation of  the effective coupling 
$G_S$ by
\be
\delta G_S = - \frac{N_c}{2\pi^2}G_S^2~\delta t.
\ee
The correction given by the figure (b) is given by
\bea
{\rm (b)} &=& 
\frac{1}{2} \left(
\frac{2G_S}{\Lambda^2}
\right)^2 \int'dp~
\bar{L}_i \frac{i \delta_l^j}{\psla}L^k ~\bar{L}_k\frac{i\delta_j^l}{\psla}L^i
+ (L \leftrightarrow R)
\nn \\
&=& - \frac{N_f}{16\pi^2}G_S^2 \delta t~\frac{1}{\Lambda^2}{\cal O}_V.
\eea
Therefore this generates the correction to $G_V$ given by 
\be
\delta G_V =  \frac{N_f}{16\pi^2}G_S^2~\delta t.
\ee
We may repeat the similar calculations, though there are 29 one-loop
diagrams involving two four-fermi couplings in total.

\begin{figure}[tb]
\begin{center}
\includegraphics[width=90mm]{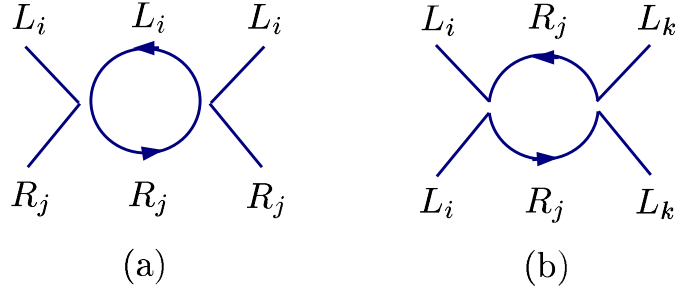}
\label{appfigab}
\caption{The typical examples of diagrams for the RG corrections to
the effective four-fermi couplings involving 2 $G_S$ interactions.}
\end{center}
\end{figure}

Another typical four-fermi diagram is given by the gauge interactions 
as shown in Fig.~18, which induces a correction of order of $g^4$.
The gauge propagator in the Landau gauge is
\be
D_{\mu \nu} = \frac{1}{p^2}\left(
g_{\mu \nu} - \frac{p_{\mu}p_{\nu}}{p^2}
\right).
\ee
The correction given by the diagram (c) is evaluated as
\bea
{\rm (c)} &=&
g^4 \sum_{A,B}~\int' dp ~ 
\bar{L}_i T^A \gamma^{\mu} \frac{i\delta^i_k}{\psla}T^B\gamma^{\rho}L^k
\bar{R}_l T^B \gamma^{\sigma} \frac{i\delta^l_j}{\psla}T^A\gamma^{\nu}R^j
D_{\mu \nu} D_{\rho \sigma} \nn \\
&=& 
\frac{9}{16}\left(N_c - \frac{2}{N_c}\right)\frac{g^4}{8\pi^2}\delta t
~\frac{1}{\Lambda^2}{\cal O}_S
-\frac{9}{32}\frac{1}{N_c^2}\frac{g^4}{8\pi^2}\delta t
~\frac{1}{\Lambda^2}{\cal O}_{V1},
\eea
where we used the identity given by (\ref{groupid}).
Therefore this correction induces the following variations,
\bea
\delta G_S &=& 
- \frac{9}{16}\left(N_c - \frac{2}{N_c}\right)\frac{g^4}{8\pi^2}\delta t,
\\
\delta G_{V1} &=&
\frac{9}{32}\frac{1}{N_c^2}\frac{g^4}{8\pi^2}\delta t.
\eea
We may find 4 different diagrams of order of $g^4$ including the crossed 
ladder diagrams and evaluate them
in a similar way.

\begin{figure}[tb]
\begin{center}
\includegraphics[width=40mm]{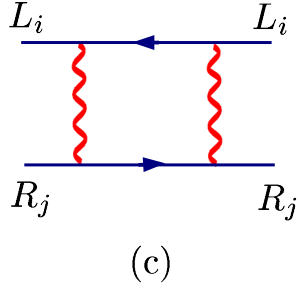}
\label{appfigc}
\caption{The typical example of diagrams for the RG corrections to
the effective four-fermi couplings involving 4 gauge interactions. }
\end{center}
\end{figure}

Moreover there are 12 diagrams proportional  $G_i g^2 (i=S, V, V1, V2)$.
Thus we may derive the RG flow equations for the effective four-fermi
couplings given by Eq.~(\ref{gsfloweq}-\ref{gv2floweq}).
It is noted that the 3-points or the 4-points interactions of the gauge
bosons do not contribute to the 1PI diagram with 4 external fermions
in the one-loop order.

\end{document}